\documentstyle[twocolumn,aps,prl,epsf]{revtex}
\newcommand{\gdot}{$\dot{\gamma}$ \/}
\newcommand{\gdotl}{$\dot{\gamma} \lambda$ \/}
\newcommand{\gdotsql}{$\dot{\gamma} \sqrt{\lambda}$ \/}
\newcommand{\gdotN}{$\dot{\gamma}_N$ \/}
\newcommand{\gdotlN}{$\dot{\gamma}_N \lambda_N$ \/}
\newcommand{\gdotsqlN}{$\dot{\gamma}_N \sqrt{\lambda}_N$ \/}
\newcommand{\lN}{$\lambda_N$ \/}

\newcommand{\beq}{\begin{equation}} 
\newcommand{\eeq}{\end{equation}}

\newcommand{\WiR}{$\dot{\gamma} \sqrt{\lambda} \tau_R$ \/}
\newcommand{\WiRN}{$\dot{\gamma}_N \sqrt{\lambda}_N \tau_R$ \/}

\begin{document} 
\twocolumn[\hsize\textwidth\columnwidth\hsize\csname
@twocolumnfalse\endcsname
\widetext
\draft

%\preprint{}

\title{Dynamics of Entangled Polymeric Fluids in Two-roll Mill studied 
via Dynamic Light Scattering and Two-color flow Birefringence. 
II. Transient flow}

\author{Subrata Sanyal\cite{byline}, Dmitry Yavich\cite{byline1} 
and L. Gary Leal} 
\address{Department of Chemical Engineering, University of California 
at Santa Barbara, Santa Barbara, CA 93106-5080, U.S.A.}

%\date{\today}
\date{June 23, 2000}

\maketitle

%\widetext

\begin{abstract}

We present an extensive experimental study of birefringence and velocity-gradient components 
for a series of high molecular weight, flexible, entangled polystyrene solutions subjected 
to transient start-up flows in a co-rotating two-mill to several Weissenberg numbers.  The 
time-dependent changes in the polymer microstructure, as measured by a two-color flow 
birefringence technique, is shown to be very distinctly coupled with the transient response 
of the corresponding velocity-field, measured by a dynamic light scattering technique.  As 
expected, polymer deformations induce substantial modifications from the Newtonian flow-field. 
As a result, measured values of both velocity-gradient components as well as flow-type 
parameter reduce, thereby drastically decreasing the extensional strength or measured 
Weissenberg number at the stagnation-point relative to the velocity-field for a Newtonian fluid.  
This result is very similar to that seen with dilute polymer solutions, although the mechanism 
responsible should be different.  Both birefringence and flow-parameters show distinct 
undulations in their transient response which appear at earlier times with increased intensity 
as the Weissenberg number of the flow is increased.  These undulations are shown to be strongly  
correlated to each other in time, indicating that the same underlying dynamics should be 
responsible for both.  The transient variation of magnitude of the measured flow-parameters is 
found to be most pronounced for the polymer solution with the smallest number of entanglements 
per chain.  Although the impact of the dynamics of both polymer on flow and flow on polymer are 
very complicated, their individual behaviors and the tight coupling between them are shown to 
be invariant to the changes in polymer concentration and molecular weight, provided that the 
number of entanglements per chain, $N_e$ is held fixed.  The dependence of the transient 
results on $N_e$ is very similar to that found in the steady-state experiments on the 
same polymeric fluids (Sanyal, Yavich, and Leal, manuscript in preparation, 2000).  The results 
presented in this paper are in close qualitative agreement with the recent theoretical 
predictions (Remmelgas and Leal, J. Non-Newtonian Fluid Mech.  {\bf 90}, 187, 2000) using a 
reptation-based vector model.  \\

\end{abstract}

%\pacs{PACS numbers: ?????}

]

\narrowtext

\section{Introduction}

Characterization of dynamics of entangled polymeric fluids subjected to macroscopic 
deformations imposed by a flow-field is of great importance for processing of polymers,  
with the objective of producing materials of desirable properties.  In most cases, this 
requires to generate large deformation-induced anisotropy in the polymer microstructure, 
i.e., a strong alignment and stretch of the polymer chains along preferential directions.  
However, most experimental efforts on entangled polymers till date were restricted to 
cases that involve relatively small conformational changes, such as in viscometric flows, 
and there have been only a few studies of entangled polymers subjected to purely 
extensional and/or extension-dominated ``mixed'' flows\cite{astarita} (also called 
``strong'' flows, since the magnitude of the strain-rate exceeds the vorticity in these 
flows), where large deformations on the polymer conformation can be achieved.  The 
primary reasons for this were, firstly, the difficulty in producing small-scale laboratory 
flows with such extensional characters, and secondly, the non-availability of simultaneous 
flow characterization procedures for such flows, compared to the case of nearly homogeneous 
viscometric flows where the flow-field does not change much from the simpler Newtonian 
form and hence stress (or birefringence) can be measured and interpreted globally within 
the flow.  An added reason for the lack of prior experiments with strong flows is that 
for complete characterization of dynamics one has to take into account of the tight coupling 
that exists between flow effects and polymer conformational effects.  That is, when large 
deformations are imposed on a polymeric fluid, the same flow properties that cause the 
conformation of polymer molecules to change, are also affected by the changes of the 
polymer molecules.  Thus, in order to produce meaningful data for the characterization of 
polymeric materials it is required to measure, simultaneously and independently, the polymer 
conformation as well as the properties of the flow-field.  

On the theoretical frontier, most efforts in recent years were focused in developing 
molecularly based constitutive equations for entangled polymer solutions.  In particular, 
the original Doi and Edwards (DE) reptation based tube model\cite{DE} was extended to 
include non-uniform chain-stretching to give rise to the so called Doi-Edwards-Marrucci-Grizzuti 
(DEMG) model\cite{DEMG,pearson}.  Although very successful in predicting polymer dynamics for 
homogeneous flows, numerical simulations of these models for non-trivial flows become too 
complex and computationally non-tractable.  In order to avoid this difficulty, many 
simplified approximations\cite{approx,vector} were suggested over the number of years, 
of which the latest one is that due to Remmelgas {et al.}\cite{vector}, which proposed a
single vector approximation to the DEMG model and separately treated the dynamics of 
stretch and orientation of the entangled polymers in flow, thereby making the numerical 
simulations of the same simpler.  

The model comparison of experimental results on entangled polymeric solution subjected to 
steady-state or time-dependent flows in the past have followed the traditional rheological path 
of focusing on shear-flows\cite{zebrowski,jim}, primarily because of the earlier-stated reason.  
On the other hand, for entangled polymeric fluids subjected to steady-state or time-dependent 
extension-dominated flows, the situation was far more complex, because even though the changes 
in the polymer conformation in such a flow could be measured with the use of a two-color flow 
birefringence (TCFB) technique\cite{fuller}, simultaneous unambiguous measurement of 
flow-parameters was not available until the advent of a dynamic light scattering 
(DLS) technique\cite{fuller,wang} in recent years.  In absence of a full-fledged fluid 
mechanical calculation for the simultaneous prediction of flow-field and birefringence in such 
flows using the DEMG model, one useful way would be to compare the experimental birefringence 
results with that obtained from the numerical simulations of the DEMG model, by using the DLS 
measurements of the velocity-gradients as input to the model.  This too have been possible only 
for steady-state flows\cite{dmitry,pst1} recently, and the work is underway to develop numerical 
codes for time-dependent flows\cite{jim1}.  In addition, the recent work of Remmelgas {et 
al.}\cite{vector} allows one to calculate prediction for the dynamics of both polymer 
microstructure and the flow simultaneously using a model which approximates the DEMG model, as 
noted earlier.  In this respect, it is extremely important to create an experimental database 
for entangled polymers subjected to extension-dominated flows which will provide a unique basis 
to test the predictions from the aforementioned reptation based models. 

With this in mind, the purpose of the present series of papers is to understand the dynamics 
of entangled, high molecular weight polymeric fluids subjected to the strong\cite{astarita}, 
extension-dominated flows generated at the stagnation-point of a co-rotating two-roll mill.  
An important characteristic of strong flows is that adjacent material points separate 
exponentially in time (and this is necessary to efficiently achieve high polymer extensions), 
rather than linearly as found in weak (shear) flows.  We have undertaken an extensive study 
of a series of entangled polystyrene solutions, where the polymer conformation is measured 
with the use of a TCFB method\cite{fuller} and the corresponding velocity-gradients of the 
flow-field are obtained by using a DLS technique\cite{fuller,wang} developed in our laboratory.  
The results from such a study are presented in the present series of papers.  In the 
Part I\cite{pst1} of this series, we have presented the results when the polymeric fluids are 
subjected to steady-state flow conditions in a two-roll mill, where a detailed comparison of 
the TCFB results for the samples with the DEMG model\cite{DEMG,pearson} is also carried out.  
In this paper, we present transient measurements of the birefringence and velocity-gradient 
in start-up of a strong, extensional-type two-roll mill flow from rest, to gain 
useful insight about the time-evolution of the coupled dynamics of the polymer configuration 
and the imposed velocity-field from an initial condition to a steady-state. 

The paper proceeds first by introducing in sec.~II the experimental methods and the samples 
used in this study.  The results from our time-dependent experiments are detailed in sec.~III, 
where we begin (in sec.~III.A) with characterizing the inception of flow with the use of a 
Newtonian fluid.  We then show how the flow-field is modified relative to that expected for a 
Newtonian fluid by the presence of polymers.  The measurements of the dynamical evolution of 
the birefringence, orientation angle, velocity-gradient and flow-type parameters following the 
start-up of a flow from rest for the three entangled samples of our study are described in 
sec.~III.B.  Section~IV deals with the discussion of the results, where we compare the flow- 
and polymer-dynamics of the samples at similar rates of flow-deformations (i.e., Weissenberg 
numbers) in sec.~IV.A, characteristic strains at the overshoot of birefringence for inception 
of flow in sec~IV.B, and the measurements made at different positions in the gap between the 
rollers of the two-roll mill in sec~IV.C.  Finally, sec.~V contains a summary of our findings 
and conclusion. 

\section{Experimental methods and materials} 

As described in Part I\cite{pst1}, we use three moderately entangled polystyrene 
solutions, PS81, PS82 and PS2, and a Newtonian fluid sample.  The Newtonian sample 
is a suspension of polystyrene microspheres of diameter $0.11 \mu$m in glycerol 
at a concentration of $150$ ppm.  As given in Table~\ref{table1}, the entangled 
polymeric fluids are made by mixing high molecular weight fairly monodisperse 
($M_w/M_n \sim 1$) polystyrene samples in solvents.   The solvents were prepared 
by adding polystyrene oligomer with $M_w = 6000$ or $2500$ and a broad molecular 
weight distribution to toluene at specific mixing ratios by weight (toluene: 2500 
PS = 43:57 for PS81 and PS82, and 48:52 for PS2).  The rheological measurements 
for simple shear flow were performed using the cone and plate geometry of Rheometrics 
RMS-800 rheometer with a cone diameter of 40 mm and a gap angle of $4^\circ$.  
The three polystyrene solutions along with their characteristic rheological 
parameters are listed in Table~\ref{table1}, which is reproduced from 
Part I\cite{pst1}.  Here, $M_w$ and $M_n$ specify the weight-average and the 
number-average molecular weights respectively, $c$ is the polystyrene concentration, 
$\eta_0$ is the zero-shear viscosity, $N_e$ is the average number of entanglements 
per chain, $\tau_R$ is the longest Rouse relaxation time, and $n_t$ is the number 
of Kuhn statistical subunits in the chain calculated from the molecular weight of 
the polymer\cite{flory} based on the assumption of 10 monomers per subunit. 

\begin{figure}
\centerline{\epsfxsize = 8cm \epsfbox{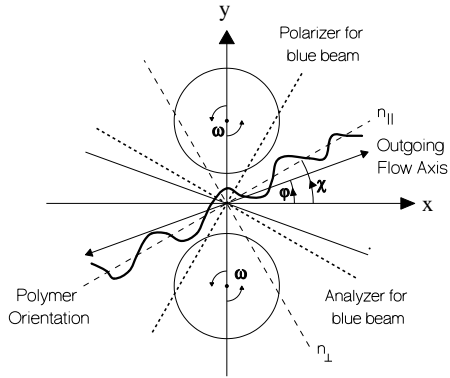}}
\caption{Relative orientation of the optical system defined by the incident blue beam 
polarization, the flow-cell, and the definition of the flow-field coordinate system.}  
\label{two-roller1}
%Fig.~1 
\end{figure}

\begin{figure}
\centerline{\epsfxsize = 8cm \epsfbox{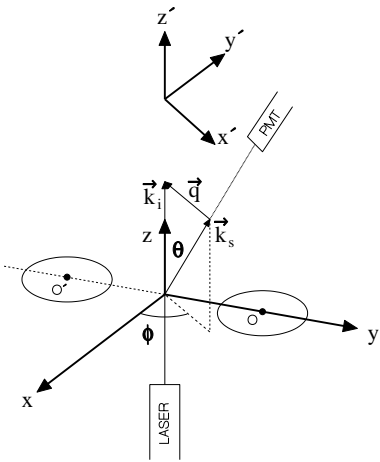}}
\caption{Schematic representation of the dynamic light scattering configuration showing 
angle definitions, vector orientations, etc.  The flow is in the $xy$-plane.}
\label{two-roller2}
%Fig.~2 
\end{figure}

Our computer-controlled co-rotating two-roll mill is a flow device\cite{enrique} 
where both rollers are driven simultaneously by a single stepping motor to generate 
strong or extension-dominated flows in the region between the rollers.  Fig.\ 
\ref{two-roller1} shows the principal optical axes of the solution and the 
symmetry axes of the flow-field, as well as the relative orientation of the polarizer 
and analyzer for the blue laser beam (used in our flow birefringence experiments, 
as discussed below).  Even though the flow is quite complex throughout the flow 
device, in our $75 \mu$m diameter measuring region surrounding the stagnation-point 
in the central plane ($xy$-plane in Figs.\ \ref{two-roller1} and \ref{two-roller2}) 
of the two-roll mill, the Newtonian velocity-field can be well approximated by a 
two-dimensional, linear (or homogeneous), symmetric form, 
\begin{equation} 
v = {\bf \nabla \vec{v}} \cdot \vec{r} = \dot{\gamma} \left[ \begin{array}{cc} 
0 & 1  \\
\lambda & 0 
\end{array} \right] \cdot \vec{r},  
\label{v} 
\end{equation}
in the coordinate system (Figs.\ \ref{two-roller1} and \ref{two-roller2}) 
that is rotated through $45^\circ$ 
relative to the principal axes of the rate of strain tensor.  Here \gdot is the 
magnitude of the velocity-gradient tensor (i.e., $\dot{\gamma} \equiv |{\bf \nabla 
\vec{v}}|$) which is a linear function of the roller rotation rate for a Newtonian 
fluid at very low Reynolds numbers, and $\lambda$ is the flow-type parameter which 
ranges from $-1$ to 1.  Pure rotational flow corresponds to $\lambda = -1$, simple 
shear flow to $\lambda = 0$ and pure extensional (or hyperbolic) flow to 
$\lambda = 1$.  With the use of different pairs of rollers from a set of eight such 
pairs, which changes the ratio of the gap-width between the rollers to their 
radius, our two-roll mill is capable of producing mixed shear and extensional flows 
with $0 < \lambda_N \le 0.25$ for a Newtonian fluid, where the subscript ``N'' used 
for any parameter throughout this paper refer to the Newtonian-value of that parameter.  
The two-roll mill configuration used in all experiments reported in the present series 
of papers\cite{pst1} correspond to \lN= 0.1501, although as we will show below the actual 
flow-type changes with the use of polymer solutions due to strong flow-modification 
compared to the flow-field expected for a Newtonian fluid.  In the stagnation region, 
the polymer has a large 
residual time and hence experiences large strain and strain-rates.  These make it 
straightforward to highly stretch and align the polymer chains at the stagnation-point, 
and owing to the substantial residence time in the region of our measurement 
the polymer will reach a configuration consistent with the corresponding homogeneous 
flow.  Thus, in the stagnation region, we expect to obtain a maximum degree of stretch 
of the polymer, and also a considerable degree of flow-modification from the Newtonian 
velocity-field.

As noted before, the two non-intrusive experimental methods used in this study are flow 
birefringence that provides a local measurement of the polymer configuration in the 
two-roll mill, and dynamic light scattering that measures the local velocity-gradients.  
The details of our entire experimental set-up comprising of these two optical techniques 
are described in Part I\cite{pst1}, and hence will not be repeated here.  

In transient flow experiments both the degree of optical anisotropy (termed the 
birefringence, $\Delta n$) and the orientation $\chi$ of the principal axes of the 
refractive index tensor relative to axes [$(x, y)$ in Figs.\ \ref{two-roller1} and 
\ref{two-roller2})] fixed 
in the flow device are unknown and varying in time.  The method of two-color flow 
birefringence\cite{fuller}, adopted in our laboratory, provides a means to 
simultaneously solve for $\Delta n$ and $\chi$ at each point in time in a single 
experiment.  In this method, the two principal spectral lines of an argon-ion laser 
(green, with the wavelength of $\lambda_G = 5145$ \AA, and blue, with $\lambda_B = 
4880$ \AA) are polarized $45^\circ$ relative to each other and then passed in a 
collinear fashion along an identical optical path through the sample in a flow-cell 
which is placed between crossed polarizers for each beam.  The extinction angle $\chi$ 
is a measure of the average orientation of the Kuhn segments in the polymer fluids.  A 
full description of the set-up and the data analysis procedure followed in the two-color 
flow birefringence experiments applied to two-roll mill flow device are previously 
detailed elsewhere\cite{enrique}. 

The complete theory of the light scattering experiments relevant to our set-up can 
be found in Refs.~\cite{fuller,wang}.  Fig.\ \ref{two-roller2} shows the orientation 
of the incoming light, $\vec{k}_i$, the scattered light, $\vec{k}_s$, and the scattering 
vector, $\vec{q} = \vec{k}_i - \vec{k}_s$.  Provided that the seed particles are 
isotropic scatterers, and the time scale associated with the velocity-gradient 
[$t_{\dot{\gamma}} \sim (q \dot{\gamma} L)^{-1}$, where $L$ is the laser beamwidth] is 
much smaller than the time scale for diffusive motion [$t_D \sim (Dq^2)^{-1}$], and the 
laser beam profile is Gaussian, the homodyne intensity auto-correlation function for a 
linear, two-dimensional flow of the form given by Eqn.\ (\ref{v}) can be related to the 
velocity-gradient tensor.  In particular, when the scattering vector $\vec{q}$ is 
oriented parallel to the $x$ axis, i.e., $\phi = 0^\circ$ [see Fig.\ \ref{two-roller2}], 
the homodyne intensity auto-correlation function reduces to 
\beq
F_2 (\vec{q}, t) = \beta \exp\left\{-\frac{1}{2} q^2 \dot{\gamma}^2 L^2 t^2
\cos^2\left(\frac{\theta}{2}\right) \right\},  
\label{F2t0}
\eeq
and when $\vec{q}$ is oriented along $y$ axis, i.e., $\phi = 90^\circ$, then 
\beq
F_2 (\vec{q}, t) = \beta \exp\left\{-\frac{1}{2} q^2 \dot{\gamma}^2 \lambda^2 
L^2 t^2 \cos^2\left(\frac{\theta}{2}\right) \right\},  
\label{F2t90}
\eeq
where $\beta$ is a coherence factor for isotropic scatterers, and the scattering angle 
$\theta$ and the flow-cell orientation angle $\phi$ define the scattered beam relative 
to the coordinates of the flow-cell, as shown in Fig.\ \ref{two-roller2}.  Thus, it 
is clear from Eqns.\ (\ref{F2t0}) and (\ref{F2t90}) that \gdot and $\lambda$ can be obtained 
from the exponential decay rates of the intensity auto-correlation functions independently 
measured at these two special orientations of the two-roll mill.

\begin{figure}
\centerline{\epsfxsize = 8cm \epsfbox{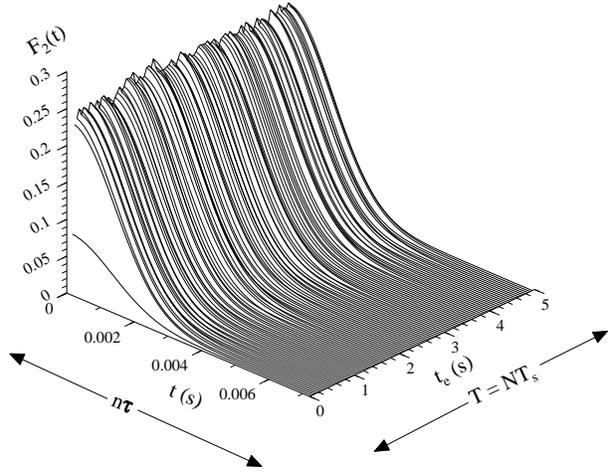}}
\caption{Typical time-resolved intensity auto-correlation functions for 
the Newtonian fluid in the startup of a steady flow, measured at the 
stagnation-point of a two-roll mill with $\lambda_N = 0.1501$, apparent 
scattering angle $\theta^\prime = 22^\circ$, roller orientation $\phi = 
0^\circ$, angular velocity of the rollers $\omega = 1.885$ rad/s.  The 
correlator sample duration and the delay time are $T_s = 0.05$~s and 
$\tau = 100 \mu$s, respectively.}
\label{3DN} 
%Fig.~3 
\end{figure}

\begin{figure}
\centerline{\epsfxsize = 8cm \epsfbox{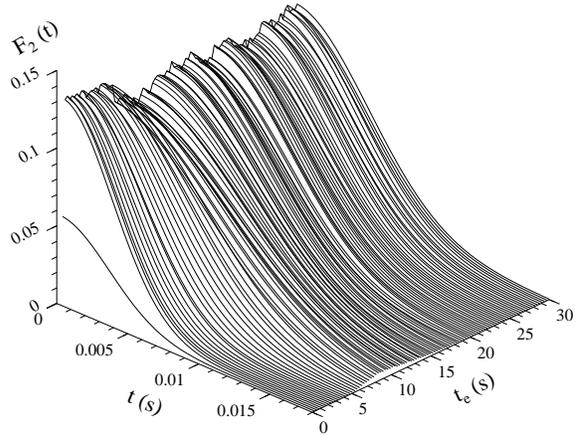}}
\caption{Typical time-resolved intensity auto-correlation functions for the polystyrene 
solution PS82 in the startup of a steady flow with $\lambda_N = 0.1501$ at the angular velocity 
of the rollers $\omega = 0.2513$ rad/s (or the measured Weissenberg number, $Wi_R = \dot{\gamma} 
\lambda^{1/2} \tau_R = 0.384$).  In every $T_s = 0.3$~s, one correlation function is registered 
at $\phi = 0^\circ$, $\theta^\prime = 19^\circ$.  The correlation delay time is $\tau = 250 \mu$s.}
\label{3Dps} 
%Fig.~4
\end{figure}

For the transient flow experiments, we use the data analysis technique described 
in Ref.\ \cite{wang}.  Here, we mention a few salient features related to these 
time-resolved experiments.  First, the correlation delay time $\tau$ should 
be so chosen that each correlation function decays completely over the $n = 72$ 
channels of the correlator, i.e., $F_2 (t = n \tau)/F_2 (t = 0) \sim 10^{-3}$.  
The predetermined total transient evolution time, $T$, of the flow to be studied is 
divided into $N$ parts, and then $N$ correlation functions are collected in this 
period of time, each for an evolution time of $T_s = T/N$ (Fig.\ \ref{3DN}).  The 
sampling time, $T_s$, over which each correlation function is collected must exceed 
the decay time, $n \tau$, of the correlation function, for the correlation function 
to be valid.  The exponential decay rate of each such valid correlation function 
can then provide the velocity-gradient component at that particular instant of the 
evolution time, $t_e$.  Second, it is desirable that the flow should not change 
substantially over the sampling time-period, $T_s$, so that the correlation function 
can be well approximated by a single exponential as in Eqns.\ (\ref{F2t0}) or 
(\ref{F2t90}), depending on the orientation of the two-roll mill.  This resolution 
of time can be obtained via the choice of $N$.  
Third, each experiment over the total evolution time, $T$, consisting of $N$ correlation 
functions (or $N$ values of velocity-gradient components) should be repeated 
several times (typically, 200 times at high motor speeds to about 800 times at 
low motor speeds for the polymer samples) to form a statistically reasonable 
correlation function for each sampling time interval, $T_s$.  The rest time 
of the motor in between two such consecutive repetition should be long enough 
so that the polymers can relax to the equilibrium state as in the beginning of 
the first experiment of the set.  Finally, it is always advisable to compare 
the long-time limit (i.e., in the limit of evolution time $t_e \rightarrow T$) 
of the transient flow-parameters with independently performed steady-state measurements, 
where the data are collected and averaged after the motor has run for a time-period 
very much longer than $T$, to check if the transient flow has actually 
reached the corresponding steady-state in the total predetermined evolution time, 
$T$.  

To exemplify these features, we present the typical three-dimensional plots of the 
time-resolved correlation functions obtained for the start-up flow for a Newtonian fluid 
(Fig.\ \ref{3DN}), as well as for a polymeric fluid, PS82 (Fig.\ \ref{3Dps}), in the 
two-roll mill.  The 
angular roller velocity and the apparent scattering angles were $\omega = 1.885$ rad/s, 
$\theta^\prime = 22^\circ$ and $\omega = 0.2513$ rad/s, $\theta^\prime = 19^\circ$, 
respectively for the Newtonian sample and PS82.  The apparent scattering angle 
$\theta^\prime$ is different from $\theta$ owing to the refraction by the sample in the 
flow-cell.  A more detailed discussion of the results obtained from these figures  
will be done in the next section.  At this point it is sufficient to note that these 
two three-dimensional plots clearly indicates the distinctly different dynamical 
evolution of the flow-field for the Newtonian and the polymeric fluid, as expected, on 
the inception of flow.  As shown in Fig.\ \ref{3Dps}, in order to obtain valid 
(exponential) time-resolved correlation functions, for polymeric fluids used in our 
experiments, that satisfy all the criteria mentioned above, we had chosen the total 
evolution time, $T \sim 30$~s (in similarity with the total evolution time for the 
corresponding transient birefringence experiments\cite{dmitry}, as well as that used 
before for other polymeric fluids in our laboratory\cite{enrique,dunlap,graham}), the 
sampling time, $T_s \sim 0.3$~s (and hence $N \sim 100$), the correlation delay time, 
$\tau = 250 \mu$s, and about 3 mins of rest time of the motor between two consecutive 
transient runs.  Thus, the total experimental time, inclusive of the repetitions 
for statistical averaging, varied from about 10 hrs. for experiments with low motor 
speeds to about 50 hrs. for that with high motor speeds.  It was shown in Part 
I\cite{pst1} that the long-time value of the flow-parameters in our $T = 30$~s transient 
experiments are within $10 \% - 15 \%$ of their value obtained from the corresponding 
steady-state experiments performed independently at the same motor speed, where the 
data collection started only after the motor has run for $\ge 2$ mins.  The results 
in Fig.\ \ref{3DN} and \ref{3Dps} are obtained by averaging over a set of 200 and 500 
repetitions, respectively. 

\begin{figure}
\centerline{\epsfxsize = 8cm \epsfbox{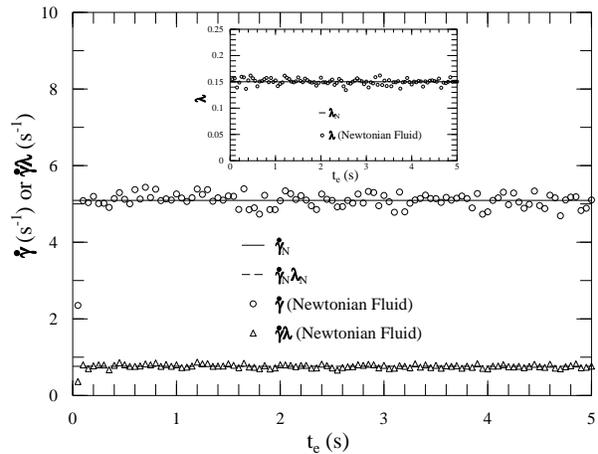}}
\caption{The magnitude of the time-dependent velocity-gradients $\dot{\gamma}$ and $\dot{\gamma} 
\lambda$ deduced from the correlation functions at $\phi = 0^\circ$ (the correlation delay time 
$\tau = 100 \mu$s), and $\phi = 90^\circ$ ($\tau = 400 \mu$s), respectively in startup of a steady 
flow for the Newtonian sample corresponding the case shown in Fig.~2.  The inset shows the 
time-dependence of the corresponding flow-type parameter $\lambda$ obtained by point to point 
division of $\dot{\gamma}\lambda$ by $\dot{\gamma}$.  The straight lines represent the creeping 
flow solutions.}
\label{gdotN} 
%Fig.~5
\end{figure}

For the experiments with the Newtonian fluid in Fig.\ \ref{3DN}, the choice of the 
correlation delay time $\tau = 100 \mu$s was made similarly to the case of polymeric 
fluids, i.e., by checking the correlation functions in the preliminary experiments to 
decay by 3 to 4 orders of magnitudes over the fixed decay time of $72 \tau < T_s$ of 
our correlator at all evolution time, $t_e$.  This sets a lower limit on $T_s$ which 
our choice of $T_s = 0.05$~s does satisfy.  The choice of $N = 100$ was made in 
similarity with Fig.\ \ref{3Dps} and hence the total evolution time in this case is 
$T = N T_s = 5$~s. 

\begin{figure}
\centerline{\epsfxsize = 8cm \epsfbox{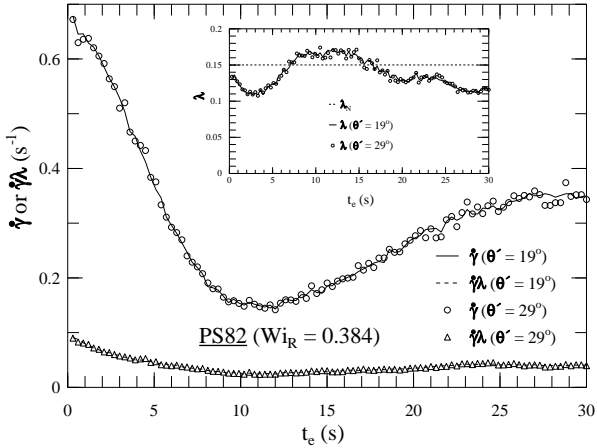}}
\caption{The magnitude of the time-dependent velocity-gradients $\dot{\gamma}$ and $\dot{\gamma} 
\lambda$ extracted from the correlation functions at $\phi = 0^\circ$ (the correlation delay time 
$\tau = 250 \mu$s) and $\phi = 90^\circ$ ($\tau = 1000 \mu$s), respectively in startup of a 
steady flow for the polystyrene sample PS82 corresponding to the case shown in Fig.~3.  Two sets 
of data, with the symbols explained in the figure, represent the results from two identical set 
of experiments at the apparent scattering angles of $\theta^\prime = 19^\circ$ and $\theta^\prime 
= 29^\circ$.  The inset shows the time-dependence of the corresponding flow-type parameter 
$\lambda$, obtained by point to point division of $\dot{\gamma} \lambda$ by $\dot{\gamma}$, 
compared to the constant $\lambda_N = 0.1501$.}
\label{gdotps} 
%Fig.~6
\end{figure}

\section{Results}

For a concentrated polymer solution, as we have mentioned before, because of the 
complicated interaction between the changes in the polymer configuration and its impact 
on the flow-field and vice versa, it is very crucial that both polymer conformation and 
flow kinematics be measured.  In this section, we will present experimental birefringence 
and dynamic light scattering results obtained for the entangled polymeric solutions 
subjected to time-dependent flows, in particular, to start-up flows from rest for the 
two-roll mill configuration which corresponds to a Newtonian flow-type parameter, \lN 
= 0.1501.  The two-color flow birefringence results, namely, birefringence, $\Delta n$, 
and orientation angle, $\chi$, versus evolution time, $t_e$, will provide information 
about the dynamical evolution of the polymer conformation following the onset of flow 
from rest.  On the other hand, the flow-parameters, namely, the velocity-gradient, 
\gdot, and the flow-type parameter, $\lambda$, plotted against evolution time, obtained 
from DLS experiments, will depict corresponding changes in the flow kinematics.  In 
Part I\cite{pst1}, the flow was shown to approximately retain its symmetry, even for 
our polymer experiments, in the range of dimensionless rates of flow-deformation (i.e., 
Weissenberg numbers) studied here.  Thus, our DLS data analysis procedure\cite{wang}, 
which is rigorously valid for seeded Newtonian fluids, can be applied for the transient 
flow experiments with polymeric fluids, presented here. 

\subsection{Flow characterization and flow-modification}

Our implementation of the two-color flow birefringence technique was primarily developed 
to study the response of polymeric fluid samples subjected to time-dependent flows, because 
of its capability of fast, point-wise measurements of the optical anisotropy of the fluid, 
with a very good reproducibility.  The dynamic light scattering technique, on the other hand, 
provides the capability of real-time, point-wise measurements of the time-dependent response 
of the flow-parameters to an imposed flow-field.  Before proceeding with the experiments on 
polystyrene solutions, it is crucial that the Newtonian flow-field at the stagnation-point 
(where the birefringence experiments are generally conducted) is known, so that one can 
quantify the changes in the measured flow-fields due to non-Newtonian behavior of the 
polymers.  It is necessary that the two-roll mill be capable of creating well-defined 
flow-field with a Newtonian fluid.  For the transient start-up flow with a Newtonian fluid, this 
would mean that the flow should instantaneously attain its steady-state value, which can be 
calculated using the theoretical creeping flow solution\cite{dunlap} (as discussed in 
Part I).  The creeping flow solutions for the flow-parameters are direct functions of the 
geometry of the two-roll mill.  In all our time-dependent flow experiments reported in this 
paper, the maximum acceleration possible with our motors (100,000 steps/s) was used to ramp 
it up from rest.  The typical picture of the evolution of the time-resolved correlation 
functions at the onset of flow in our flow characterization experiments with the Newtonian 
fluid is shown in Fig.\ \ref{3DN}.  The detailed nature of the time-resolved correlation 
functions for both parallel ($\phi = 0^\circ$) and perpendicular ($\phi = 90^\circ$) 
orientations of the two-roll mill are quite similar and hence we showed only one of them, 
$\phi = 0^\circ$, here.  The velocity-gradient components, \gdot and \gdotl, at each 
instant of evolution time, $t_e$, is extracted from the decay rates of the fitted 
exponential functions of Eqn.\ (\ref{F2t0}) and (\ref{F2t90}) to these measured 
correlation functions.  The value of the laser beamwidth, $L$, used here is obtained from 
a calibration procedure that is described in Part I.  In order to verify that the 
experimentally obtained correlation functions are very close to exponential in nature so 
that the above procedure is indeed justified, we required a correlation coefficient $R^2$ 
specifying the quality of fit exceeding 0.99.  The extracted velocity-gradient components 
from Fig.\ \ref{3DN} are shown in Fig.\ \ref{gdotN}.  We see that except for the first 
point at $T_s = 0.05$~s, both \gdot and \gdotl have reached their steady-state value, which 
is maintained throughout the experiment.  In fact, from the second point onwards, all data 
for these two parameters lie within $11 \%$ of the theoretical creeping flow 
solutions\cite{pst1,dunlap}, \gdotN and \gdotlN, shown by the straight lines.  Also, it is 
clearly shown in the inset that the time-evolution of the flow-type, $\lambda$, obtained 
by dividing \gdotl by \gdot at each evolution time, is very close to its theoretical value 
\lN = 0.1501.  The initial delay in the onset of flow, as noted in the earlier work\cite{wang}, 
is due to the finite time-scale for vorticity diffusion in the Newtonian fluid.  Thus, we 
have clearly shown that the two-roll mill is capable of creating a constant well-characterized 
flow-field in the stagnation-point.  Further, the start-up is almost instantaneous and limited 
solely by the ramp-up acceleration of the motor and the time-scale associated with diffusion 
of vorticity in the fluid.

To see how \gdot, \gdotl and $\lambda$ evolve with time in the case of a non-Newtonian 
viscoelastic sample, let us consider the representative case of the sample PS82 subjected 
to the start-up of a flow shown in Fig.\ \ref{3Dps}, and extract the flow-parameters in 
an exactly similar fashion as discussed above.  Fig.\ \ref{gdotps} presents the result, 
deduced from Fig.\ \ref{3Dps}, where as shown by the solid and broken lines respectively, 
following the start-up, both \gdot and \gdotl decrease in time to reach a minimum and then 
again increase to reach their approximate steady-state value (which is lower than the initial 
value) at a longer evolution time.  The inset shows the time-dependence of $\lambda$ as 
obtained by dividing \gdotl by \gdot, point by point.  Owing to a much higher viscosity of 
polystyrene samples compared to the Newtonian fluid, the vorticity diffusion time-scale in 
this case is comparatively much smaller and hence the initial value of \gdot is very close to 
the Newtonian-value.  The initial value of the second correlation function in Fig.\ \ref{3Dps} 
is different from the rest, reflecting a different value of the coherence factor, $\beta$, 
although its decay rate similar, thus 
yielding a similar value of \gdot.  The presence of polymer is clearly making an impact 
upon the flow-type parameter (and hence on \gdotl) from the very beginning of the 
flow-evolution.  The parameter $\lambda$ follows a complex time-evolution and shows an 
oscillation around the corresponding Newtonian-value of \lN = 0.1501.  Its value is lower 
than \lN almost always, except when both \gdot and \gdotl show minimum between $t_e = 7$~s 
and $t_e = 15$~s.  These curves, when compared to the corresponding ones for the Newtonian 
fluid (Fig.\ \ref{gdotN}), provide a clear indication that the entangled polystyrene samples 
very significantly modify the otherwise Newtonian flow.  In Fig.\ \ref{gdotps}, we have 
also provided results obtained by repeating the same experiment when the apparent scattering 
angle $\theta^\prime$ is changed from $19^\circ$ to $29^\circ$.  The time-profile for each 
flow-parameter is very well reproduced within an experimental error limit of $9 \%$.  This 
provides a direct proof of the fact that the transient flow conditions with these entangled 
samples are quite repeatable and the data obtained in the light scattering experiments for 
such flows are well reproducible inside the range of apparent scattering angle ($19^\circ 
\le \theta^\prime \le 29^\circ$) used in the present study.

\begin{figure}
\centerline{\epsfxsize = 8cm \epsfbox{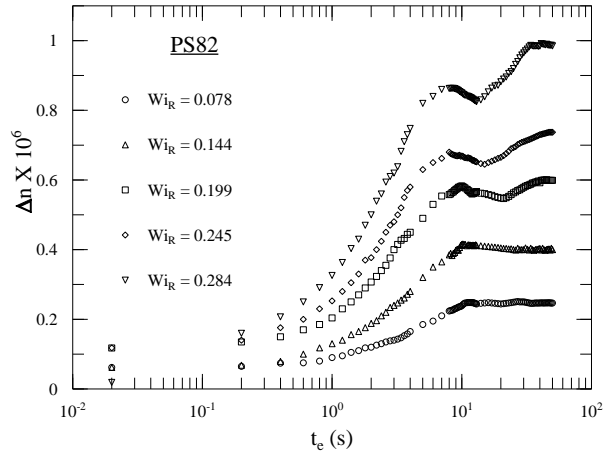}}
\caption{Time-dependent birefringence on inception of a two-roll mill 
flow ($\lambda_N = 0.1501$) for PS82 at several Weissenberg numbers, 
$Wi_R = \dot{\gamma} \lambda^{1/2} \tau_R$, 
based on the measured steady-state 
values of $\dot{\gamma}$ and $\lambda$, and the Rouse time, $\tau_R$.}
\label{n82a} 
%Fig.~7
\end{figure}

\begin{figure}
\centerline{\epsfxsize = 8cm \epsfbox{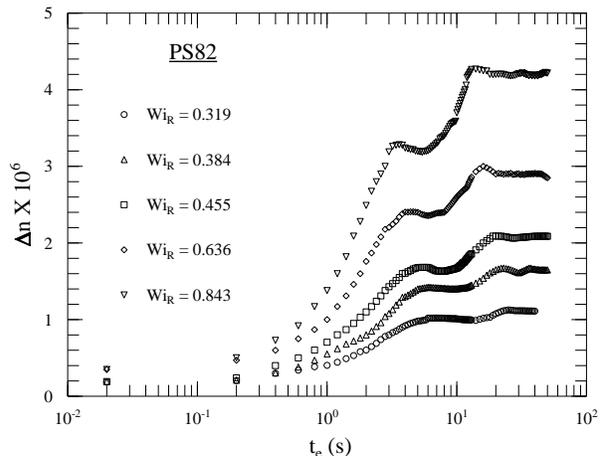}}
\caption{Same as Fig.~7, but at higher Weissenberg numbers, 
$Wi_R = \dot{\gamma} \lambda^{1/2} \tau_R$.} 
\label{n82b} 
%Fig.~8
\end{figure}

In prior work from this laboratory\cite{jim,dmitry}, the non-dimensionalized measure used 
to identify transient flow experiments was the Weissenberg number of the flow, $(Wi_R)_N = 
\dot{\gamma}_N \sqrt{\lambda}_N \tau_R$, based on the Newtonian-values of the velocity-gradient, 
$\dot{\gamma}_N$, and the flow-type parameter, $\lambda_N$.  This was done because of the 
unambiguous relationship 
between \gdotN and the roller speed $\omega$ given by the creeping flow solution\cite{dunlap}.  
In reality, however, the Weissenberg number, $Wi_R =$ \WiR, based on the measured values of 
\gdot and $\lambda$, is expected to be different for different polymeric fluids even if the 
imposed roller speed is the same.  The velocity-gradient parameters measured in the 
independently performed steady-state experiments (as discussed before) are used to calculate 
the actual Weissenberg number, $Wi_R$, which are in turn used to identify our transient 
experiments.  Table~\ref{table2} lists the Newtonian and the measured values of the 
Weissenberg numbers used in our study for three polystyrene samples.  The decreased value 
of the measured $Wi_R$ compared to its Newtonian-value, $(Wi_R)_N$, is obviously a direct 
consequence of the flow-modification.  The percentage difference of $Wi_R$ relative to 
$(Wi_R)_N$, also given in Table~\ref{table2}, shows that the flow-modification is very 
substantial in all the cases studied.  There is a decreasing trend with the increase in 
the rate of deformation for the two polystyrene fluids with the similar number of 
entanglements, PS81 and PS2, and an increasing trend with PS82.  We note that the range of 
Weissenberg numbers chosen for each of the three polystyrene samples are very different when 
compared to each other.  This is so because, firstly, the choice of roller speeds (in rad/s) 
achievable in our two-roll mill set-up is restricted with the choice of motor (i.e., the 
motor speeds available in steps/s) and the worm-gear assembly attached between the motor 
and the two-roll mill.  Secondly, based on our preliminary calculations of the number of 
entanglements per chain, $N_e$, for three samples, we did not expect the longest Rouse 
relaxation time, $\tau_R$, would be so different for the sample PS2 (see, Table~\ref{table1}) 
compared to the other two.  The linear viscoelastic experiments\cite{pst1} to extract $\tau_R$ 
of these fluids were performed only after the TCFB experiments.  Finally, the flow-modification 
was much stronger than we expected.  This, as 
can be seen from Table~\ref{table2}, has further reduced the values of the measured Weissenberg 
numbers.  Thus, even though our initial motor speeds (in steps/s) were very similar for all 
samples, the final measured Weissenberg numbers are very different.   We also note that the 
strain-rate \gdotsql (used in calculating $Wi_R =$ \WiR) provides a measure of the rate of 
stretch of material points along the direction of the outflow axis of the flow of 
Eqn.\ (\ref{v}) [see, Fig.\ \ref{two-roller1}] from the stagnation-point.  This means that 
the $Wi_R$ used here or the $(Wi_R)_N$ used in earlier reports\cite{jim,dmitry}, provides 
the steady-state value of the deformation rate which can be achieved in transient 
experiments only in the long-time limit, when the polymer chain is finally oriented along 
the principal eigenvector direction.  At earlier times, when the flow is evolving with time, 
the appropriate measure of the non-dimensionalized rate of deformation should be the 
{\em transient} Weissenberg number, $Wi_R^{tr} = \dot{\gamma} (1 + \lambda) \tau_R$, based 
on the transient strain-rate $\dot{\gamma} (1 + \lambda)$, instead of its steady-state value, 
\gdotsql. 

\begin{figure}
\centerline{\epsfxsize = 8cm \epsfbox{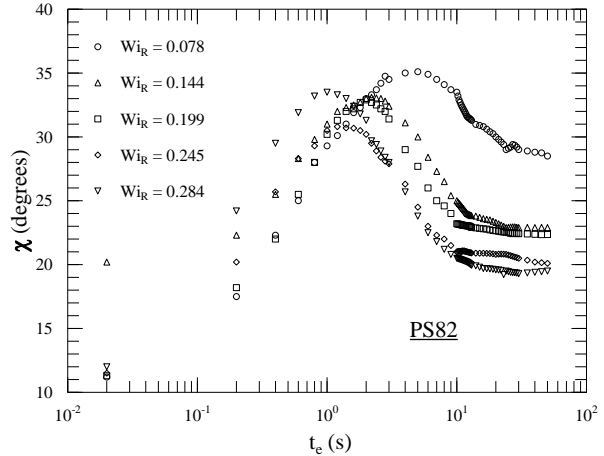}}
\caption{Flow-induced extinction angle, $\chi$ versus evolution time $t_e$ for PS82 
at different values of measured Weissenberg numbers.}
\label{chi82a} 
%Fig.~9
\end{figure}

\begin{figure}
\centerline{\epsfxsize = 8cm \epsfbox{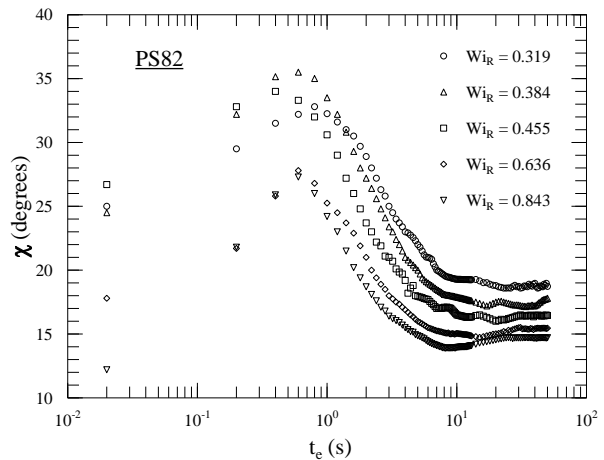}}
\caption{Same as Fig.~9, but at higher measured Weissenberg numbers, 
$Wi_R = \dot{\gamma} \lambda^{1/2} \tau_R$.} 
\label{chi82b} 
%Fig.~10
\end{figure}

\subsection{Dynamic evolution of birefringence and flow} 

As noted earlier, both the birefringence and the components of the velocity-gradients were 
measured in a region surrounding the stagnation-point of the flow-cell midway between the two 
rollers, where polymer chains can highly stretch and align.  Also, because of the substantial 
residence time in this region, the polymer will reach a configuration consistent with the 
corresponding homogeneous flow.  The complicated dynamics of the evolution of polymer 
configuration and flow-field, and their tight coupling require that we show simultaneously 
the measured birefringence, orientation angle and the corresponding flow-parameters plotted 
against evolution time, $t_e$,  for the three entangled polystyrene fluids, at several different 
Weissenberg numbers.  

We will first look into the features of the birefringence and flow data in general terms 
for each polystyrene fluids.  A more detailed discussion and a comparison of results amongst 
the samples will be done in a later section.  Let us first consider the case of PS82, the 
solution with about 7 entanglements per chain (see, Table~\ref{table1}).  Figs.\ \ref{n82a} 
and \ref{n82b} show the temporal-evolution of birefringence on inception of flow to a total of 
ten measured Weissenberg numbers spaning a range from 0.078 to 0.843 [$0.099 \le (Wi_R)_N \le 
1.782$].  The intrinsic difficulty associated with our transient birefringence experiment, 
namely, the appearance of multiple orders in the retardance, makes interpretation of results 
at high deformation rates extremely difficult and thus limits the highest value of the 
Weissenberg number for which the experiment could be performed.  At the smallest $Wi_R$, the 
birefringence increases almost monotonically until a steady-value is reached.  At higher values 
of $Wi_R$, the temporal-evolution of birefringence displays a series of undulations.  Before 
reaching the long-time value, the birefringence trace goes through a notch at $Wi_R = 0.144$.  
At higher $Wi_R$, this notch gives rise to a very distinct peak (or overshoot), which progressively 
appears at earlier times with increased magnitude and intensity as $Wi_R$ is increased.  We 
note that the peak-time, $t_p$, of the first overshoot of birefringence in PS82 have reduced 
from $t_p \approx 10$~s at $Wi_R = 0.199$ to $t_p \approx 3.55$~s at $Wi_R = 0.843$.  Also, 
the long-time value of the birefringence monotonically increases with the increase in the rates  
of flow-deformation.  These typical features are in agreement the observations in earlier studies with 
concentrated polymeric fluids for the inception of both shear flows\cite{zebrowski,jim} and 
extension-dominated flows\cite{enrique1} at large enough rates of deformation.  At still 
higher $Wi_R$, this overshoot is followed by a very distinct trough (or undershoot) and then 
a second overshoot, each of which appear at an earlier time and become more distinct with 
increasing $Wi_R$.  These features too are consistent with the prior observation by Geffroy 
{\em et al.}\cite{enrique1}, but compared to their results the dynamical evolution of the 
birefringence in this case has quite a few dissimilarities: the magnitude of the second overshoot 
and the long-time value of birefringence here is significantly higher than the magnitude of 
the first overshoot. 

\begin{figure}
\centerline{\epsfxsize = 8cm \epsfbox{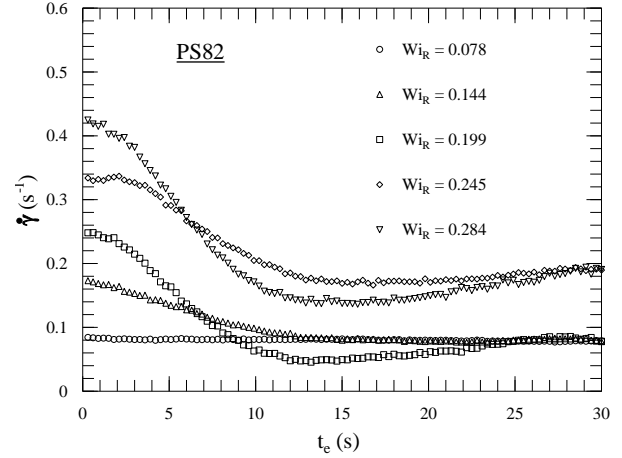}}
\caption{Plot of the velocity-gradient component, $\dot{\gamma}$ (deduced from the correlation 
functions at the two-roll mill orientation of $\phi = 0^\circ$), as a function of evolution time, 
$t_e$, for start-up experiments with the sample PS82 at different measured steady-state $Wi_R$.}
\label{g82a} 
%Fig.~11
\end{figure}

\begin{figure}
\centerline{\epsfxsize = 8cm \epsfbox{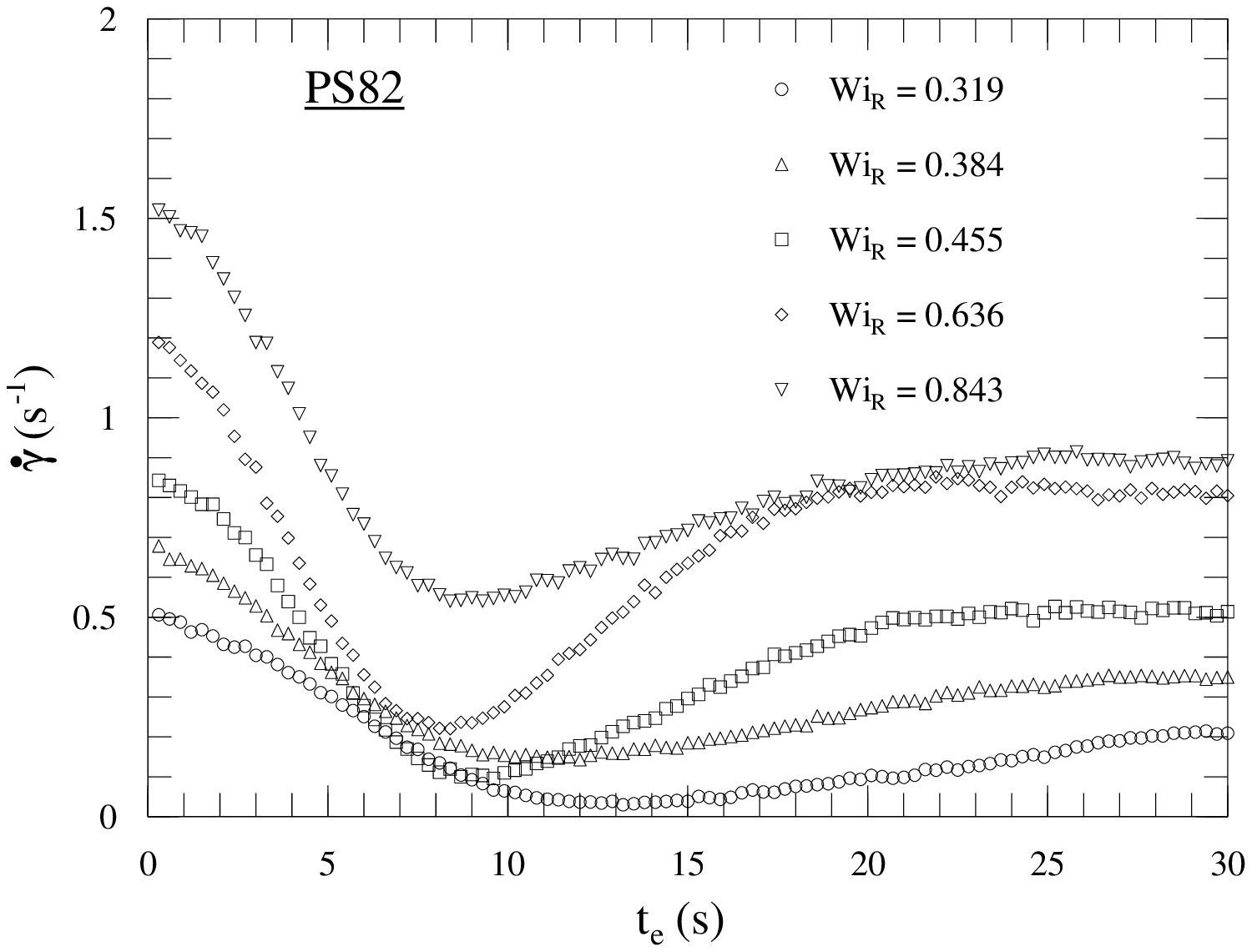}}
\caption{Same as Fig.~11, but at higher measured Weissenberg numbers, 
$Wi_R = \dot{\gamma} \lambda^{1/2} \tau_R$.} 
\label{g82b} 
%Fig.~12
\end{figure}

\begin{figure}
\centerline{\epsfxsize = 8cm \epsfbox{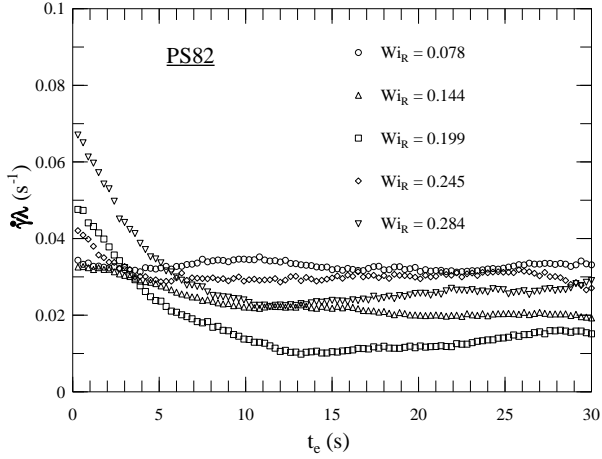}}
\caption{Time-dependent velocity-gradient component, $\dot{\gamma}\lambda$, 
for PS82, extracted from correlation functions in start-up of steady flows at 
the two-roll mill orientation of $\phi = 90^\circ$, and at different measured 
Weissenberg numbers.}
\label{gl82a} 
%Fig.~13
\end{figure}

\begin{figure}
\centerline{\epsfxsize = 8cm \epsfbox{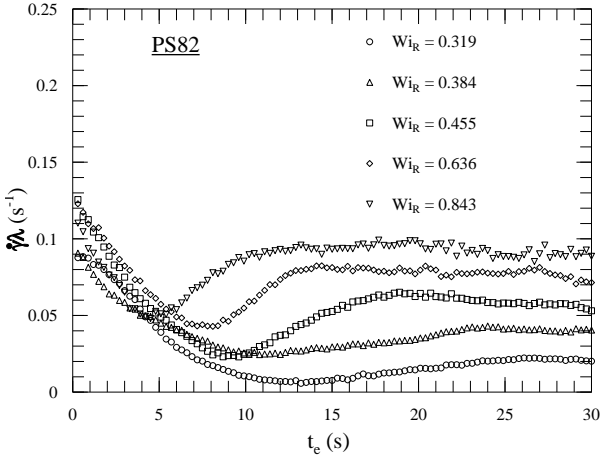}}
\caption{Same as Fig.~13, but at higher values of the measured Weissenberg numbers.}
\label{gl82b} 
%Fig.~14
\end{figure}

The corresponding orientation angle data are shown in Figs.\ \ref{chi82a} and \ref{chi82b}.  
For all cases, we should expect that the orientation angle, $\chi$,  should initially have a value 
of $45^\circ$, i.e., it should coincide with the principal axis of the rate of strain 
tensor [Fig.\ \ref{two-roller1}].  In our experiments, the orientation angle at the onset 
of the flow is dominated by the residual anisotropy of the glass in the flow-cell window.  The 
orientation angle reaches a maximum value in time as the anisotropic contribution of the 
polymer becomes dominant.  Keeping with the earlier studies for start-up of both 
shear\cite{zebrowski,jim} and extension-dominated flows\cite{enrique1}, 
at low Weissenberg numbers, the orientation angle almost monotonically decreases to its 
long-time value, but at high enough $Wi_R$, it shows an undershoot just after the first 
overshoot in birefringence takes place.  At large values of $Wi_R$, we may expect that 
the polymer molecules will become oriented close to the outflow axis of the flow-field.  
For a Newtonian fluid, the corresponding asymptotic value of $\chi$ is $\chi(Wi_R \rightarrow 
\infty) = \chi_\infty = \tan^{-1} (\sqrt{\lambda_N}) = 21.2^\circ$.  For the polymeric fluids, 
we may still try to find the values of $\chi_\infty$ with the use of the above relation with 
\lN replaced by the asymptotic value, $\lambda_\infty$, of the measured flow-type parameter, 
so that the flow-modification is taken into account.  This could be done since, as shown in 
Part I, the flow approximately retains its symmetry in our experiments with entangled fluids.  
As expected, $\chi_\infty$ in Figs.\ \ref{chi82a} and \ref{chi82b} decreases faster with the 
increase in Weissenberg number at smaller Weissenberg numbers, but this decrease almost saturate 
at higher Weissenberg numbers.  At the highest $Wi_R$, $\chi_\infty \sim 14.8^\circ$.  The 
orientation angle curves at intermediate values of $Wi_R$ also show some undulations before 
reaching the final asymptotic 
values, which are well-correlated in time to the first undershoot and the second overshoot in 
the corresponding birefringence traces.

\begin{figure}
\centerline{\epsfxsize = 8cm \epsfbox{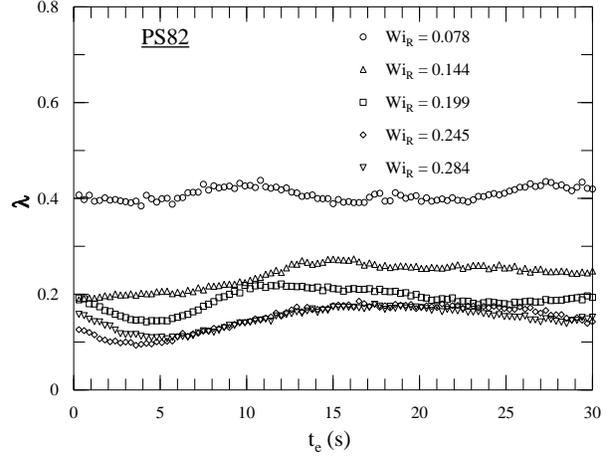}}
\caption{Inception of the flow-type parameter, $\lambda$, for PS82, deduced from the ratio of 
the square root of the exponential decay rates of two corresponding correlation functions at 
different Weissenberg numbers, $Wi_R = \dot{\gamma} \lambda^{1/2} \tau_R$ (based on the 
measured steady-state values of $\dot{\gamma}$ and $\lambda$).}
\label{l82a} 
%Fig.~15
\end{figure}

\begin{figure}
\centerline{\epsfxsize = 8cm \epsfbox{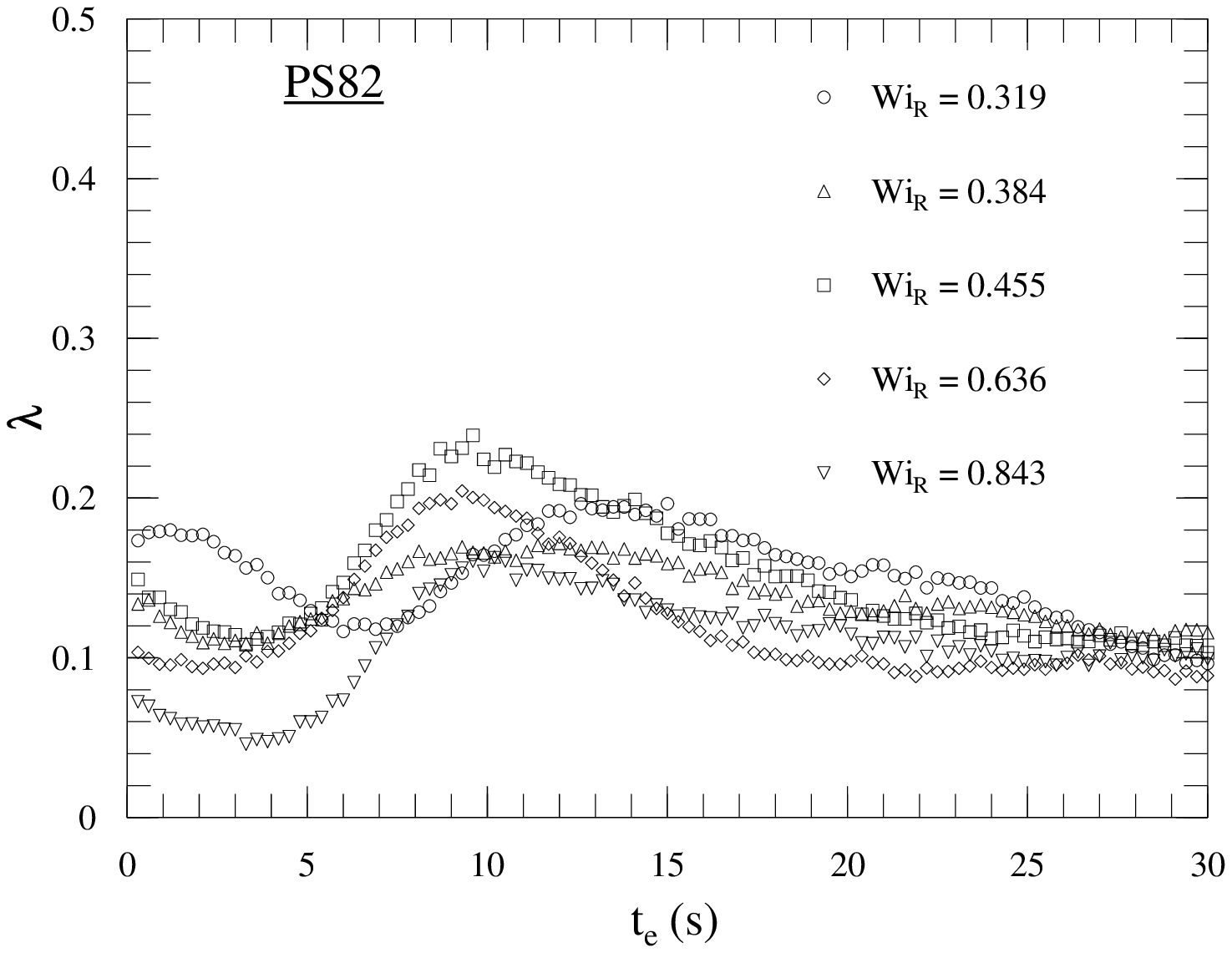}}
\caption{Same as Fig.~15, but at higher values of the measured Weissenberg numbers.}
\label{l82b} 
%Fig.~16
\end{figure}

\begin{figure}
\centerline{\epsfxsize = 8cm \epsfbox{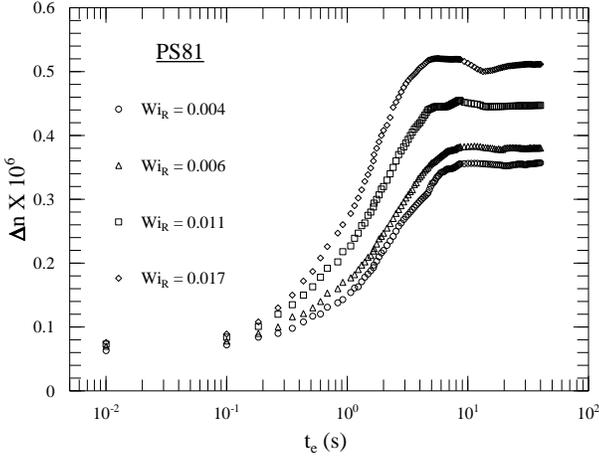}}
\caption{Inception of birefringence as a function of evolution time, $t_e$, 
for sample PS81 at different measured steady-state Weissenberg numbers, 
$Wi_R = \dot{\gamma} \lambda^{1/2} \tau_R$.} 
\label{n81a} 
%Fig.~17
\end{figure}

In Figs.\ \ref{g82a}, \ref{g82b} and Figs.\ \ref{gl82a}, \ref{gl82b}, respectively, we show 
the measured \gdot and \gdotl versus the evolution time $t_e$ for PS82 at the Weissenberg 
numbers for which we have shown the birefringence results above.  From these figures, we 
extract the dynamical evolution curves for the flow-type parameter, $\lambda$, as shown in 
Figs.\ \ref{l82a} and \ref{l82b}.  At the lowest $Wi_R$, \gdot does not show any appreciable 
change from its Newtonian-value over the entire evolution time, but with the increase of 
$Wi_R$, there is a strong deviation in time from the Newtonian flow.  The curves for \gdot ``start'' 
from the Newtonian-value (which are higher at higher $Wi_R$, as expected), pass through a 
minimum (or undershoot) and then again increase to the respective long-time values, which are 
lower than the initial Newtonian-values at the same $Wi_R$.  The 
minimum of the undershoot oscillates around $t_e \sim 14$~s for $Wi_R \le 0.319$, but 
beyond that, it progressively appears at earlier time until $Wi_R = 0.636$.  At this 
deformation rate and above, the minimum of the undershoot in \gdot appears at $t_e \sim 8$~s.  
The strength of the undershoot, as well as the asymptotic value of \gdot increases with the 
Weissenberg number.  At the two lowest $Wi_R$, \gdotl (Fig.\ \ref{gl82a} and \ref{gl82b}) and 
hence $\lambda$ (Fig.\ \ref{l82a} and \ref{l82b}) show very little variation in their values 
following the onset of the flow, but surprisingly, these values are higher than that expected 
with a Newtonian fluid.  As we will show and discuss about it in the following, this behavior 
takes place repeatably at low Weissenberg number flows for all three entangled solutions 
studied.  For the velocity-gradient component, \gdotl, too the intensity of the undershoot 
increases with increasing rate of deformation and also it progressively appears at earlier times.  
We have noted that the initial magnitude of the velocity-gradient component \gdot is always 
very close but slightly less than the Newtonian value, \gdotN.  In contrast, the initial 
value of \gdotl for higher rates of deformation is such that the initial value for $\lambda$ 
oscillates around the \lN and then reduces substantially from the same at the highest 
Weissenberg numbers.  At low rates of flow-deformation, the flow-type parameter shows an 
undershoot at $t_e \sim 5$~s before reaching the long-time value by $t_e \sim 15$~s, but at 
high rates, apart from the presence of this undershoot, an overshoot shows up at $t_e \sim 
10$~s before the flow-type parameter reaches the asymptotic limit by $t_e \sim 25$~s.  There 
is a transition between these two types of behaviors at $Wi_R = 0.319$, where the extremum 
of both the undershoot and the following overshoot are shifted to the later times.  The 
asymptotic value, $\lambda_\infty$, of the flow-type parameter also shows a difference in 
behavior below and above this $Wi_R$.  In particular, $\lambda_\infty$ decreases as $Wi_R$ 
is increased till $Wi_R = 0.319$, but for higher values of $Wi_R$, it remains almost 
constant, $\lambda_\infty \sim 0.1$.  

\begin{figure}
\centerline{\epsfxsize = 8cm \epsfbox{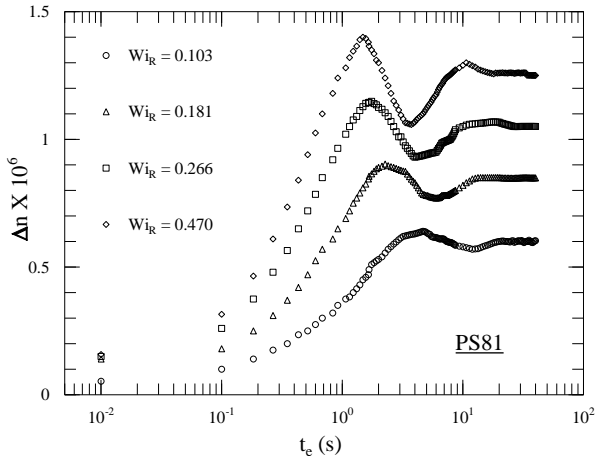}}
\caption{Same as Fig.~17, but at higher Weissenberg numbers.}
\label{n81b} 
%Fig.~18
\end{figure}

\begin{figure}
\centerline{\epsfxsize = 8cm \epsfbox{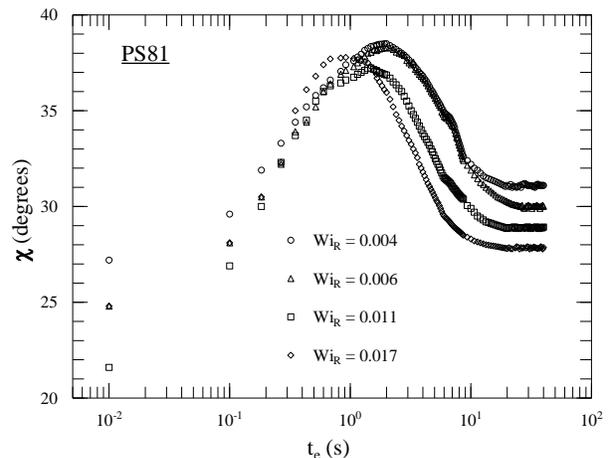}}
\caption{Time-dependent orientation angle on inception of a two-roll mill 
flow for PS81 at several different values of measured steady-state $Wi_R$.}
\label{chi81a} 
%Fig.~19
\end{figure}

\begin{figure}
\centerline{\epsfxsize = 8cm \epsfbox{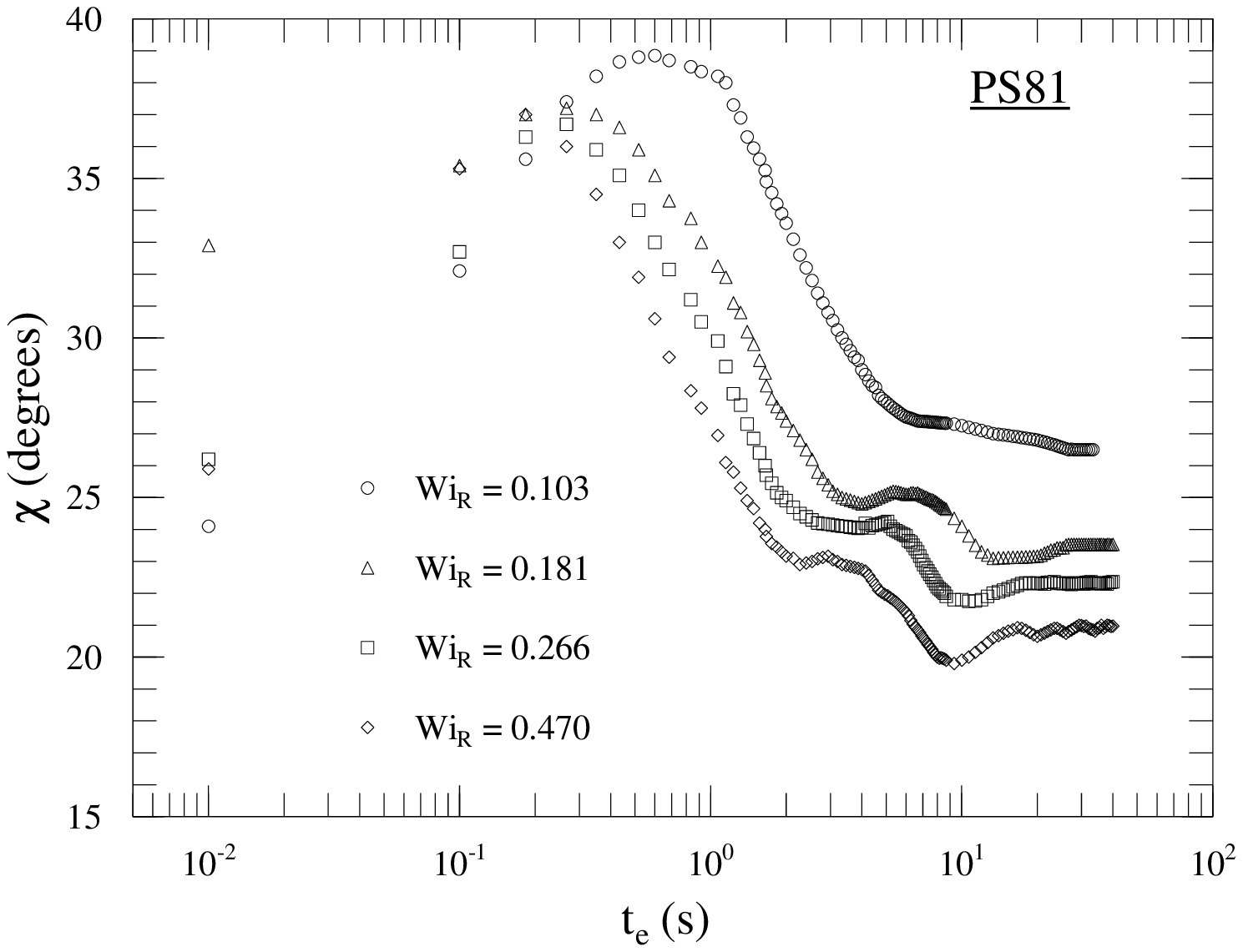}}
\caption{Similar to Fig.~19, but at higher measured rates of flow-deformation, 
$Wi_R = \dot{\gamma} \lambda^{1/2} \tau_R$.} 
\label{chi81b} 
%Fig.~20
\end{figure}

In Figs.\ \ref{n81a}, \ref{n81b} and Figs.\ \ref{chi81a}, \ref{chi81b}, we show the 
temporal-evolution of the birefringence and orientation angle, respectively, for the start-up 
of the two-roll mill flow to several different Weissenberg numbers with the polystyrene fluid 
PS81, which has approximately 13 entanglements per chain.  At low Weissenberg numbers, similar 
to the results shown above for sample PS82, both birefringence and orientation angle (after 
overcoming the residual glass anisotropy) show smooth transitions towards their long-time 
values.  On the other hand, at high rates of deformation, the time-dependent configurational 
dynamics for PS81 is very much different compared to that of PS82, as apparent from these figures.  
The overshoot and the following undershoot in the birefringence curves become considerably 
sharper with the increase of $Wi_R$ and a second overshoot in the birefringence is clearly 
apparent only at the highest $Wi_R$.  With the increase of $Wi_R$, the polymer response time 
reduces (as can be seen from the shift of the temporal positions of these overshoots and 
undershoots toward shorter times) and the asymptotic birefringence monotonically decreases.  
Despite these similarities, in sharp contrast to the observations made with PS82, the 
long-time value of the transient birefringence in PS81 is always lower than the maximum in 
the first overshoot.  Also, at high values of $Wi_R$, the orientation angle shows a distinct 
undershoot, followed by an overshoot and a second undershoot, which too become more intense 
with increasing $Wi_R$.  There is a strong correlation in their temporal positions with that 
of the overshoot, undershoot and a second overshoot found in the birefringence data at the 
same $Wi_R$.  Interestingly, the second undershoot in the orientation angle is clearly visible 
at a lower $Wi_R$ before the second overshoot appears in the corresponding birefringence.  At 
the highest $Wi_R$, the data for $\chi$ versus $t_e$ shows some ringing before reaching its 
long-time value, $\chi_\infty$.  In similarity with the sample PS82, here too the 
conformational dynamics appear different below and above a transition Weissenberg number 
($Wi_R = 0.103$, for PS81).  At $Wi_R > 0.103$, the extremum in the undulations in both 
birefringence and orientation angle move faster to earlier times, and $\chi_\infty$ 
approaches faster to $\chi_\infty \sim 21^\circ$ at the highest $Wi_R$.  

\begin{figure}
\centerline{\epsfxsize = 8cm \epsfbox{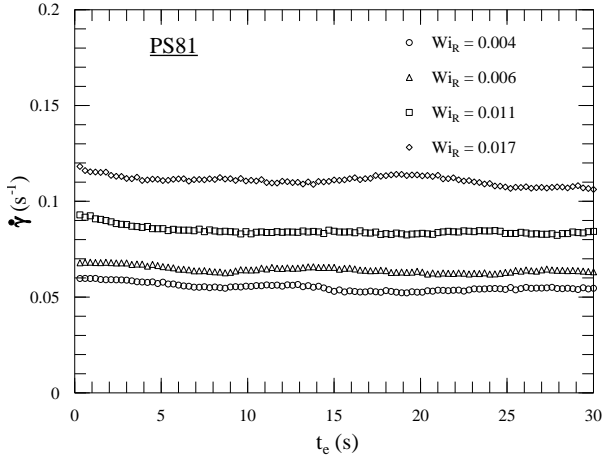}}
\caption{Flow-induced changes in the evolution of velocity-gradient component, 
$\dot{\gamma}$, obtained from the exponential decay rate of the correlation 
functions at $\phi = 0^\circ$ in start-up flow for PS81 and different measured 
steady-state Weissenberg numbers.}
\label{g81a} 
%Fig.~21
\end{figure}

\begin{figure}
\centerline{\epsfxsize = 8cm \epsfbox{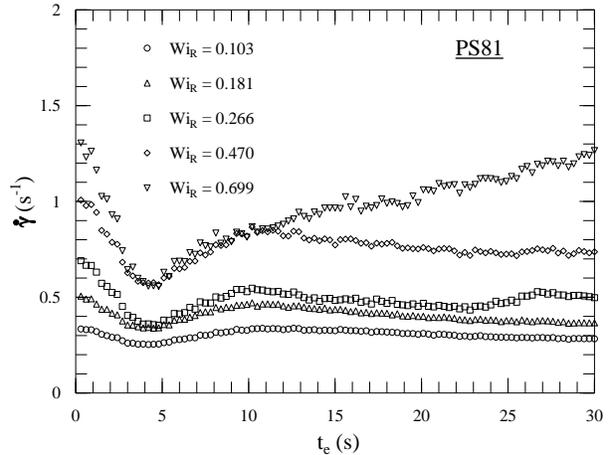}}
\caption{Similar to Fig.~21, but for higher values of $Wi_R$.}
\label{g81b} 
%Fig.~22
\end{figure}

\begin{figure}
\centerline{\epsfxsize = 8cm \epsfbox{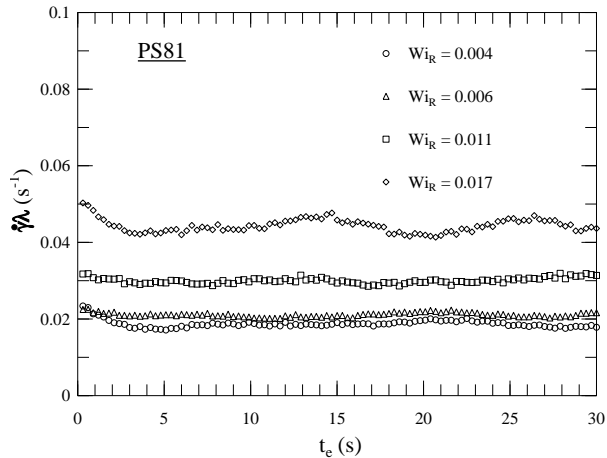}}
\caption{Transient velocity-gradient component, $\dot{\gamma}\lambda$, for 
PS81, versus evolution time, $t_e$, for inception of a two-roll mill flow at 
several measured steady-state 
$Wi_R = \dot{\gamma} \lambda^{1/2} \tau_R$.} 
\label{gl81a} 
%Fig.~23
\end{figure}

\begin{figure}
\centerline{\epsfxsize = 8cm \epsfbox{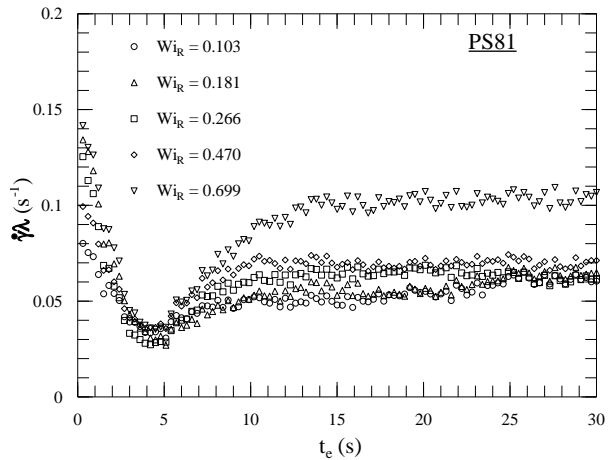}}
\caption{Same as in Fig.~23, but at higher $Wi_R$.}
\label{gl81b} 
%Fig.~24
\end{figure}

\begin{figure}
\centerline{\epsfxsize = 8cm \epsfbox{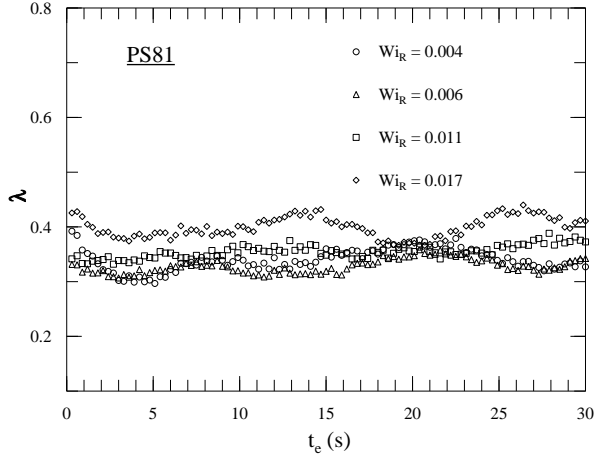}}
\caption{Time-dependence of the flow-type parameter, $\lambda$, obtained via 
point-by-point division of the corresponding data for $\dot{\gamma}\lambda$ by 
$\dot{\gamma}$ for PS81 at different measured steady values of $Wi_R$.} 
\label{l81a} 
%Fig.~25
\end{figure}

The dynamic evolution of the velocity-gradient components, \gdot and \gdotl for the sample 
PS81 at the Weissenberg numbers corresponding to Figs.\ \ref{n81a} to \ref{chi81b} are 
shown in Figs.\ \ref{g81a} to \ref{gl81b}, respectively.  The flow-type parameter versus 
the evolution time, $t_e$, at the same values of $Wi_R$, extracted from these two results 
are depicted in Figs.\ \ref{l81a} and \ref{l81b}.  At the lowest rates of deformation, both 
measured components of the velocity-gradient, \gdot (Fig.\ \ref{g81a}) and \gdotl (Fig.\ 
\ref{gl81a}), (and hence $\lambda$ in Fig.\ \ref{l81a}) remain almost constant over the entire 
evolution time of $T = 30$~s, although there is a tendency of a slight undershoot at $t_e \sim 
4$~s in \gdotl curves.  The impact of the changes in polymer configuration at these different 
Weissenberg numbers is more apparent on \gdotl and hence on $\lambda$, compared to the 
corresponding \gdot.  This is clearly seen from these figures where \gdot approximately retains 
the Newtonian value, \gdotN, in time, but \gdotl and $\lambda$ are higher than \gdotlN and \lN, 
respectively.  At higher $Wi_R$, both \gdot and \gdotl show an undershoot behavior at $t_e \sim 
4$~s before reaching their corresponding long-time values by $t_e \sim 15$~s, except for the 
case of the highest deformation rate, $Wi_R = 0.699$, where \gdot continues to increase 
(Fig.\ \ref{g81b}) and the curve for \gdotl has a higher asymptotic value than the rest of the 
curves in Fig.\ \ref{gl81b}.  The initial value of the flow-type parameter reduces from about 0.4 
(Fig.\ \ref{l81a}) to about 0.1 (Fig.\ \ref{l81b}) with the increase of $Wi_R$, primarily because 
of the corresponding changes seen in the initial value of $\dot{\gamma}\lambda$.  In contrast, 
we have noted that for both PS82 and PS81, \gdot always ``begins'' with a value very close to the 
Newtonian-value, $\dot{\gamma}_N$.  As seen for \gdot and \gdotl versus $t_e$ curves, the flow-type 
parameter for PS81 too undershoots at $t_e \sim 4$~s.   In similarity with the observations made 
for PS82, the asymptotic value of $\lambda$ at the highest Weissenberg number reduces to 
$\lambda_\infty \sim 0.1$, and there is a a clear-cut difference in the temporal behavior of the 
flow-parameters below and above a transition rate of flow-deformation, $Wi_R = 0.103$.  Compared 
to PS82, the overall variation of magnitude of the flow-parameters over the total evolution time 
is less pronounced for PS81. 

\begin{figure}
\centerline{\epsfxsize = 8cm \epsfbox{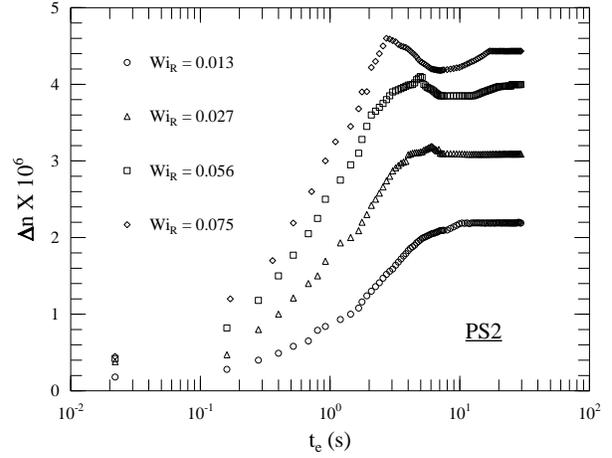}}
\caption{Birefringence for PS2 solution in start-up flows at different measured 
dimensionless rates of flow-deformation, 
$Wi_R = \dot{\gamma} \lambda^{1/2} \tau_R$.} 
\label{n2a} 
%Fig.~27
\end{figure}

\begin{figure}
\centerline{\epsfxsize = 8cm \epsfbox{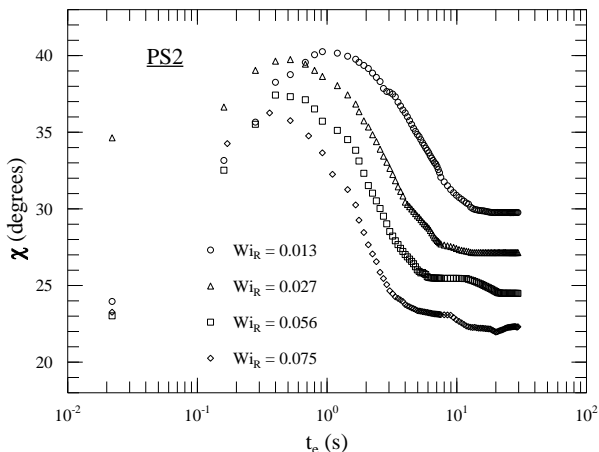}}
\caption{Plot of the dynamic evolution of orientation angle on inception of steady-flows for PS2 
at several steady-state values of measured $Wi_R$.}
\label{chi2a} 
%Fig.~29
\end{figure}

In general terms, the results for the solution PS2 (having about 13 entanglements per chain) 
shown in Figs.\ \ref{n2a} to \ref{chi2b}, is very similar to PS81 (also having $N_e \sim 13$).  
In particular, the feature of the monotonic increase of birefringence to its asymptotic value 
at lowest $Wi_R$ changes to that of the appearance of an overshoot, followed by an undershoot 
and another overshoot with the increased rate of deformation.  In similarity with our 
observations on the other two samples, here too all these undulations seem to appear earlier in 
time and their amplitudes increase with $Wi_R$.  In contrast to PS82 and PS81, at highest rates 
of deformation, the asymptotic birefringence for PS2 in Fig.\ \ref{n2b} is similar to the value 
at the first overshoot.  In this case too, there is a critical Weissenberg number, $Wi_R = 
0.075$, below which the evolution in both $\Delta n$ and $\chi$ is monotonic but above this 
number, they show undulations which become stronger as the Weissenberg number is augmented.  Each 
undershoot (or overshoot) in the orientation angle curve is strongly correlated and a little 
shifted in time compared to the corresponding overshoot (or undershoot) in the birefringence 
curve.  With the increase of $Wi_R$, here too the asymptotic birefringence gradually increase 
and the asymptotic orientation angle, $\chi_\infty$, gradually decrease.  At the highest $Wi_R$, 
$\chi_\infty \sim 18.5^\circ$, in Fig.\ \ref{chi2b}.  The prime difference in this case, compared 
to PS81, is that the magnitude of birefringence at the second overshoot is similar to the first 
overshoot and the asymptotic value.  

\begin{figure}
\centerline{\epsfxsize = 8cm \epsfbox{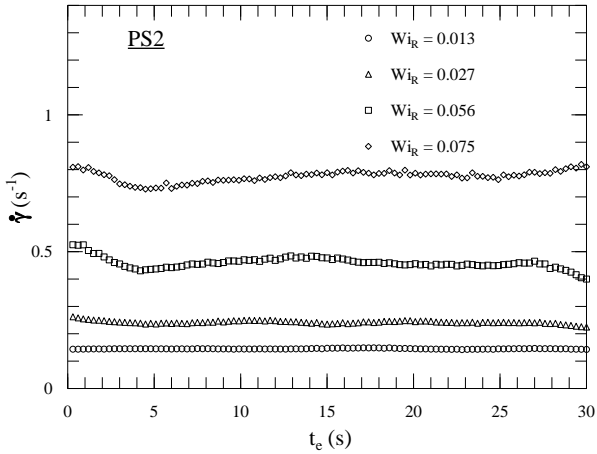}}
\caption{Transient $\dot{\gamma}$ versus $t_e$ in start-up of steady-flows 
for PS2 at different measured $Wi_R$.}
\label{g2a} 
%Fig.~31
\end{figure}

\begin{figure}
\centerline{\epsfxsize = 8cm \epsfbox{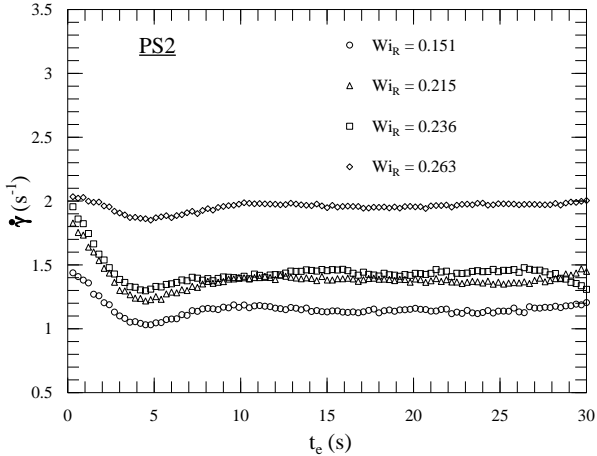}}
\caption{Same as Fig.~31, but at higher measured values 
$Wi_R = \dot{\gamma} \lambda^{1/2} \tau_R$.} 
\label{g2b} 
%Fig.~32
\end{figure}

\begin{figure}
\centerline{\epsfxsize = 8cm \epsfbox{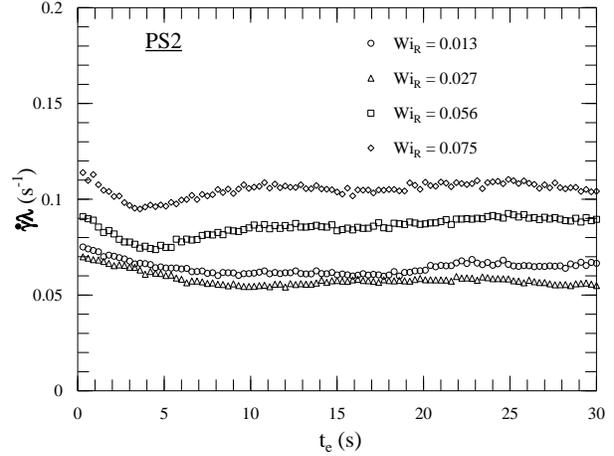}}
\caption{Transient velocity-gradient component, $\dot{\gamma}\lambda$, versus $t_e$ 
for PS2 with the two-roll mill orientation $\phi = 90^\circ$, and at different $Wi_R$.}
\label{gl2a} 
%Fig.~33
\end{figure}

\begin{figure}
\centerline{\epsfxsize = 8cm \epsfbox{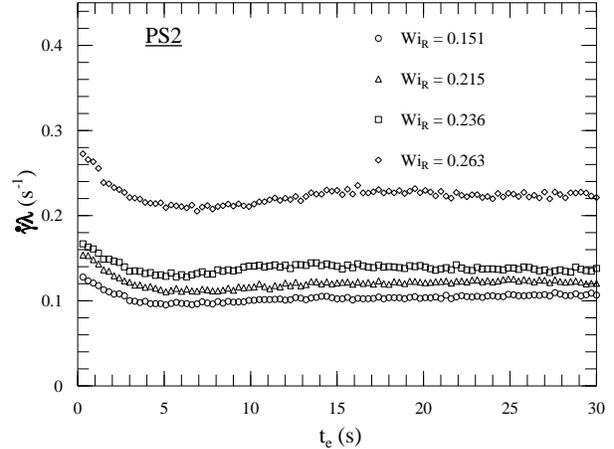}}
\caption{Same as Fig.~32, but at higher measured values of Weissenberg numbers.}
\label{gl2b} 
%Fig.~34
\end{figure}

\begin{figure}
\centerline{\epsfxsize = 8cm \epsfbox{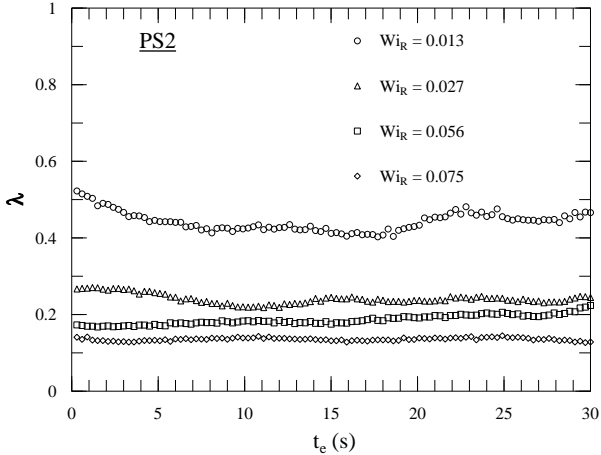}}
\caption{Flow-induced changes in flow-type parameter, $\lambda$, versus $t_e$ for PS2 
at different measured rates of flow-deformations, 
$Wi_R = \dot{\gamma} \lambda^{1/2} \tau_R$.} 
\label{l2a} 
%Fig.~35
\end{figure}

\begin{figure}
\centerline{\epsfxsize = 8cm \epsfbox{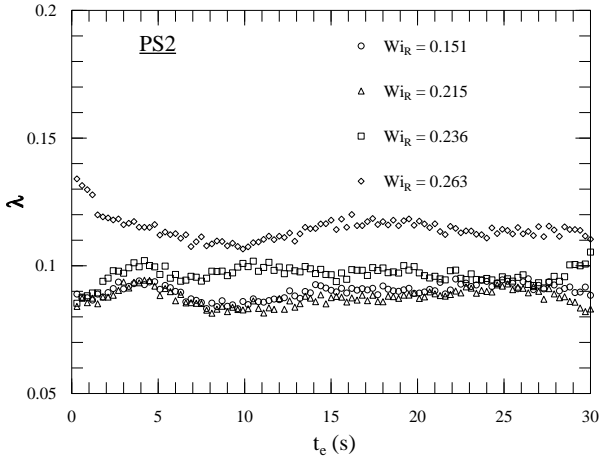}}
\caption{Same as Fig.~35, but at higher measured $Wi_R$.}
\label{l2b} 
%Fig.~36
\end{figure}

As shown in Figs.\ \ref{g2a} to \ref{l2b}, the characteristic evolution of the measured 
flow-parameters for PS2 is, again, extremely similar to that for PS81 (which has similar $N_e 
\sim 13$), but is very different from that seen in PS82 ($N_e \sim 7$).  Here too \gdotl and 
hence $\lambda$ seems to be much more sensitive than \gdot to the changes in the polymer 
configuration in that their initial values are always different than that expected with a 
Newtonian fluid, which is not the case with \gdot.  Similar to PS81, the polymer-induced variation 
in the magnitude of flow-parameters for PS2 over the entire period of evolution time is not as 
pronounced as seen for PS82.  The small undershoots at $t_e \sim 4$~s present in \gdotl versus 
$t_e$ curves at all Weissenberg numbers and also in \gdot versus $t_e$ curves at intermediate and 
high $Wi_R$ almost cancel each other in $\lambda$ versus $t_e$ results, which are derived from 
the division of these two data sets.  If at all, there is a little overshoot present in $\lambda$ 
versus $t_e$ curves at the similar evolution time $t_e \sim 4$~s for $0.151 \le Wi_R \le 
0.236$.  The asymptotic limit of the flow-parameters are reached much earlier than $t_e \sim 
15$~s.  Since the overall variation of the flow-parameters over the entire evolution time is not 
too pronounced for PS2 for any Weissenberg number, the changes in their transient evolution below 
and above the transition Weissenberg number, $Wi_R = 0.075$, is not too appreciably noticed here, 
except for the case of flow-type parameters in Figs.\ \ref{l2a} and \ref{l2b}. 

\section{Discussion of results}

\begin{figure}
\centerline{\epsfxsize = 8cm \epsfbox{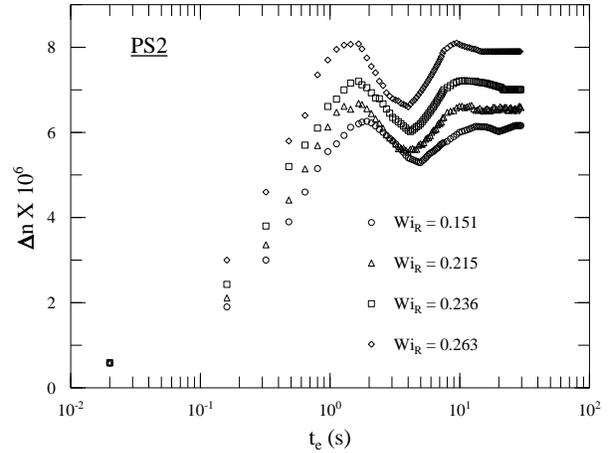}}
\caption{Same as Fig.~27, but at higher measured 
$Wi_R = \dot{\gamma} \lambda^{1/2} \tau_R$.} 
\label{n2b} 
%Fig.~28
\end{figure}

Turning back to the results for the three entangled samples, the prime important feature to 
note is that the dynamical response of the polymer systems, PS81 and PS2, with a similar 
number of entanglements per chain appears to be very much similar, even though they have 
quite different concentrations, $c$, and molecular weights, $M_w$ (see, Table~\ref{table1}).  
On the other hand, the dynamics is very different from these two for the sample PS82, which has 
a different number entanglements per chain.  This should be expected, based upon the scaling 
ideas associated with reptation modelling, provided we stay away from chain-stretching.  In the 
limit of high $Wi_R$, the DEMG version\cite{DEMG,pearson} of the reptation model, used in Part I, 
predicts a saturation of the dimensionless steady-state birefringence of $\sim n_t/N_e$ 
corresponding to a maximum chain-extension ratio of $\sim \sqrt{n_t/N_e}$.  As in Part I, we 
first non-dimensionalize the asymptotic values of the birefringence shown in Figs.\ \ref{n82b}, 
\ref{n81b}, and \ref{n2b} at the highest $Wi_R$ by scaling them with the birefringence $C G_N^0$ 
(obtained by using the stress-optical law) that would be present at a stress level equal to the 
plateau modulus, $G_N^0$.  The values used for the stress-optical coefficient, $C$, and the 
plateau modulus are same as in Part I\cite{pst1}.  Comparing these values to the DEMG predictions 
given above, we see that the experimentally observed maximum asymptotic chain-extension for these 
samples are $\sim 44.04 \%$ for PS82, $\sim 6.63 \%$ for PS81, and $\sim 11.04 \%$ for PS2.  This 
is a clear indication of the fact that we are quite away from the chain-stretching behavior.  The 
impact of the changes in the polymer conformational dynamics, as depicted by the time-evolution 
of the measured components of velocity-gradients, $\dot{\gamma}$, and \gdotl, and hence the 
flow-type parameter, $\lambda$, too is also extremely similar for PS81 and PS2, and this dynamical 
behavior is very different from that seen in the case of PS82.  

As we have mentioned before, the occurrence of an overshoot in the birefringence and an 
associated undershoot in the orientation angle at a slightly shifted time, have been observed 
earlier for the start-up of a simple shear flow ($\lambda = 0$)\cite{zebrowski,jim}, as well 
as for the start-up flow with a little more extensional character ($\lambda = 0.019$) in a 
two-roll mill\cite{enrique1}.  For simple shear flows, the onset of the overshoot in the 
first normal stress difference (birefringence) occurs at $\dot{\gamma}_N \tau_R \sim 0.4 - 
0.7$\cite{pearson,M&G}.  The onset of the birefringence overshoot in our experiments occur at 
the measured Weissenberg numbers that are very different from sample to sample, namely, at 
$Wi_R \sim 0.144$ for PS82, $Wi_R \sim 0.017$ for PS81, and at $Wi_R \sim 0.027$ for 
PS2.  These numbers correspond to $\dot{\gamma}_N \tau_R \sim 0.5$ for the less entangled 
sample, PS82, and $\dot{\gamma}_N \tau_R \sim 0.3$ for the other two more entangled samples, 
PS81 and PS2, which are in the same ballpark with the earlier results\cite{pearson,enrique1,M&G}.  

\begin{figure}
\centerline{\epsfxsize = 8cm \epsfbox{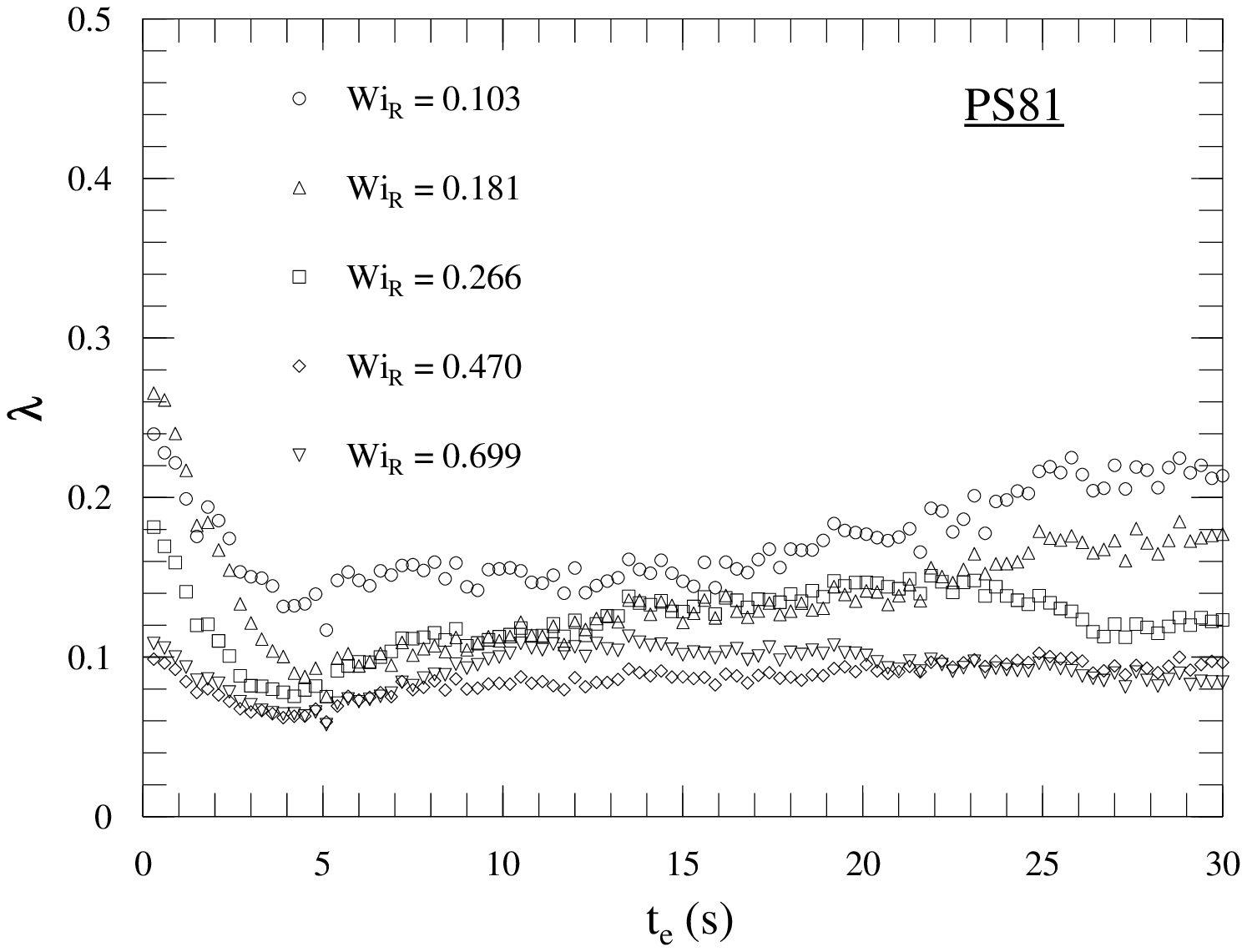}}
\caption{Similar to Fig.~25, but at higher measured rates of deformation, 
$Wi_R = \dot{\gamma} \lambda^{1/2} \tau_R$.} 
\label{l81b} 
%Fig.~26
\end{figure}

\begin{figure}
\centerline{\epsfxsize = 8cm \epsfbox{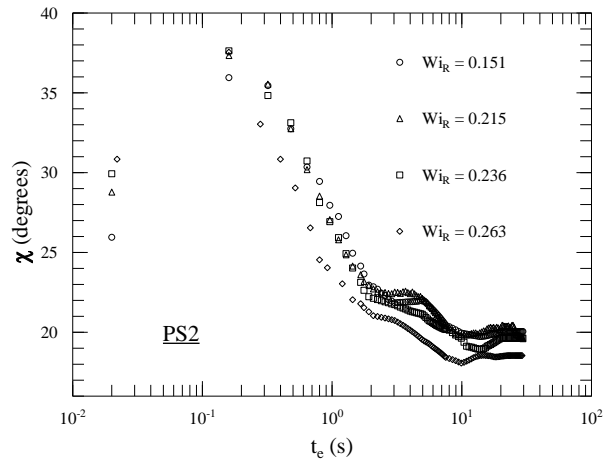}}
\caption{Same as Fig.~29, but at higher 
$Wi_R = \dot{\gamma} \lambda^{1/2} \tau_R$.} 
\label{chi2b} 
%Fig.~30
\end{figure}

Here, too the number of entanglements per chain seem to have an important role to play.  
We note that the birefringence overshoot for the inception of a simple shear flow have also 
been predicted using the Doi-Edwards reptation model with chain-stretching\cite{pearson}, 
although the undershoot in the orientation angle was not predicted.  In this work, as well as 
in prior work\cite{zebrowski,jim,dmitry,enrique1} on polymers, the undershoot behavior 
in the orientation angle seem to be very much correlated in time with the overshoot behavior 
in the birefringence.  According to the model, the transient nature of chain-stretching in a 
simple shear is responsible for the overshoots.  On the basis of the DEMG model\cite{DEMG,pearson}, 
it was speculated in earlier work\cite{jim,dmitry} from our laboratory that ``tube-dilation'' 
may be responsible for such an effect.  In simple terms, this means that a flow-deformation 
induced decrease in the number of entanglements per chain (and hence the dilation in the 
tube-radius) takes place on a slower time-scale than over which the polymer orients.  This 
faster orientation would allow polymer to over-orient (and thereby show an undershoot in the 
transient orientation angle) and then relax to its asymptotic value as $N_e$ is decreased.  
From Figs.\ \ref{l82b}, \ref{l81b} and \ref{l2b}, at the highest Weissenberg numbers, we get 
$\lambda_\infty \sim 0.10$ for PS82 and PS81, and $\lambda_\infty \sim 0.11$ for PS2.  These 
estimate $\chi_\infty \sim 17.5^\circ$ for PS82 and PS81, and $\chi_\infty \sim 18.3^\circ$ 
for PS2, but the experimental results at the highest $Wi_R$, as can be seen in Figs.\ 
\ref{chi82b}, \ref{chi81b} and \ref{chi2b} are $\chi_\infty \sim 14.8^\circ$ (PS82), 
$\chi_\infty \sim 21^\circ$ (PS81), and $\chi_\infty \sim 18.5^\circ$ (PS2), respectively.

The velocity-gradient components measured via dynamic light scattering experiments show that 
there is a significant amount of flow-modification by the polymers relative to the otherwise 
Newtonian-value.  The flow-modification appears to be very strong for all polystyrene samples 
(see, Table~\ref{table2}), but the evolution of the flow-parameters over the total duration $T$ 
of our transient experiments show most dramatic changes for the sample with the lowest 
entanglements per chain.  In comparison, this effect is rather weak for PS81 and PS2, both 
with $N_e \sim 13$.  We have noted that the velocity-gradient component, \gdotl, for the 
perpendicular orientation of the two-roll mill and hence the flow-type parameter, $\lambda$, is 
much more sensitive to the polymer induced changes in the flow than the other velocity-gradient 
component, $\dot{\gamma}$.  At the lowest $Wi_R$, the values of the flow-type parameters for all three 
samples, deduced from the corresponding \gdotl and \gdot measurements, are consistently higher 
than that expected for a Newtonian fluid.  This finding is also consistent with that in our 
steady-state experiments\cite{pst1} on the same polymeric fluids.  We note that an increase in 
the value of flow-type parameter in the vicinity of the stagnation-point of the two-roll mill, 
relative to the Newtonian-value at small $Wi_R$ was not observed by Wang {\em et al.}\cite{wang} 
because we explored the range of measured $Wi_R$ much lower than their study.  This is the only 
earlier experiment, that we are aware of, which has performed simultaneous TCFB and DLS 
measurements on entangled polymers subjected to inception of two-roll mill flows.  On the other 
hand, recent numerical simulations, using a vector approximation\cite{vector} of the DEMG reptation 
model, by Remmelgas {\em et al.}\cite{johan} indeed predicted the same result for entangled 
polymeric fluids in the same configuration (\lN = 0.1501) of the two-roll mill.  Ongoing experiments 
and theoretical work in our laboratory are intended to look into this important issue. 

For all cases studied here, as the flow evolve to high $Wi_R$, the measured values of both the 
velocity-gradient, \gdot, and the flow-type parameter, $\lambda$, and hence the strain-rate, 
\gdotsql, have reduced relative to the velocity-field for a Newtonian fluid.  The reduction of 
the extensional strength of the flow as well as the strain-rate in the neighborhood of the 
stagnation-point was observed for dilute polymers\cite{graham,ng} too for which it has been 
shown theoretically\cite{graham,singh}, using nonlinear dumbbell models\cite{CR}, to be related to the 
strain-rate hardening\cite{orr} of the extensional viscosity of these fluids.  Obviously, a 
completely different methodology has to be responsible for the entangled polymeric fluids, 
since they show a strain-rate softening of the extensional viscosity.  The recent work of 
Remmelgas {\em et al.}\cite{johan} have, in fact, predicted the decrease in the rate of strain 
for entangled polymeric fluids at the stagnation-point, in similarity with our experiments.  
This decrease was shown to result from the effect of shear thinning viscosity in the flow close 
to the rollers making the momentum transfer to the region between the rollers less efficient. 

\subsection{Comparison of dynamics at similar $Wi_R$}

\begin{figure}
\centerline{\epsfxsize = 8cm \epsfbox{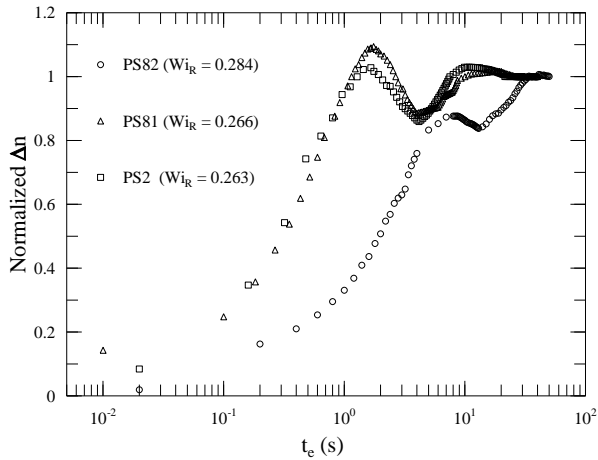}}
\caption{Plot of the time-evolution of birefringence, $\Delta n$, normalized with the 
corresponding asymptotic values, for all three polystyrene samples on inception of steady-flows at 
similar measured steady-state $Wi_R$.}
\label{samewin} 
%Fig.~37
\end{figure}

\begin{figure}
\centerline{\epsfxsize = 8cm \epsfbox{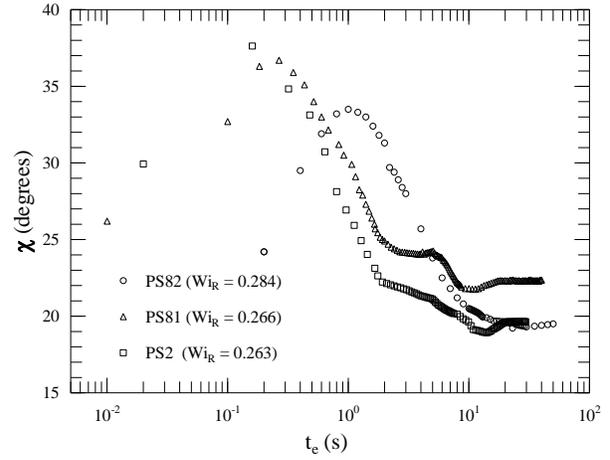}}
\caption{Same as Fig.~37, but for the orientation angle, $\chi$.}
\label{samewichi} 
%Fig.~38
\end{figure}

\begin{figure}
\centerline{\epsfxsize = 8cm \epsfbox{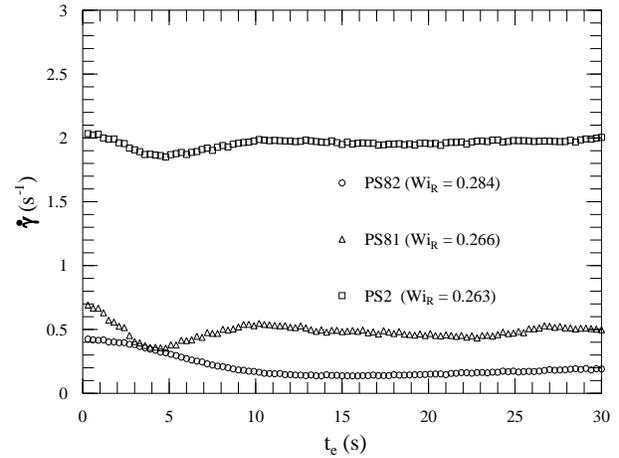}}
\caption{Transient evolution of the velocity-gradient, $\dot{\gamma}$, for all three 
polystyrene samples at similar measured $Wi_R$.}
\label{samewig} 
%Fig.~39
\end{figure}

\begin{figure}
\centerline{\epsfxsize = 8cm \epsfbox{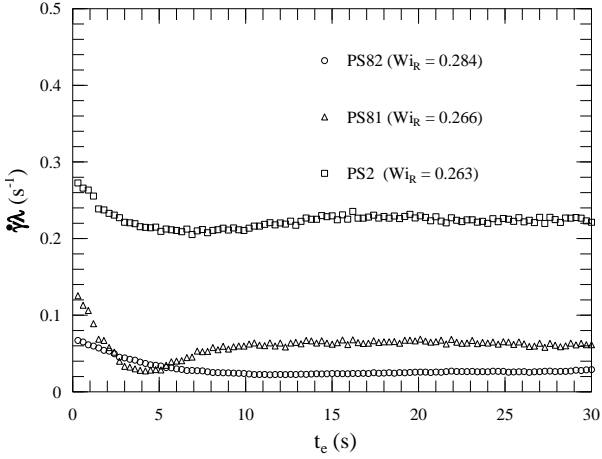}}
\caption{Same as Fig.~39, but for a different velocity-gradient component, 
$\dot{\gamma}\lambda$.}
\label{samewigl} 
%Fig.~40
\end{figure}

\begin{figure}
\centerline{\epsfxsize = 8cm \epsfbox{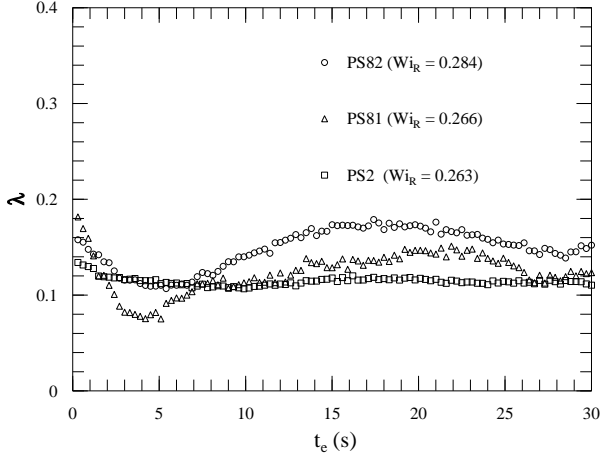}}
\caption{Dynamic evolution of the flow-type parameter, $\lambda$, on inception of 
steady-flows for all three polymeric fluids at similar measured $Wi_R$.}
\label{samewil} 
%Fig.~41
\end{figure}

We will now compare the results for the three entangled solutions at the similar values 
of the measured Weissenberg numbers, namely, $Wi_R = 0.284$, $0.266$, and $0.263$ 
for PS82, PS81 and PS2, respectively.  Fig.\ \ref{samewin} shows the birefringence traces 
versus $t_e$, where the birefringence for each sample is normalized by the asymptotic 
value, for the purpose of comparison.  The corresponding curves for the evolution of the 
orientation angle, $\chi$, and the flow-parameters, \gdot, \gdotl and $\lambda$ are shown 
in Figs.\ \ref{samewichi} to \ref{samewil}.  

One of the prime important results of this study, 
that the evolution dynamics of the polymer microstructure as well as its impact on the 
flow-field is very much dependent on the number of entanglements per chain for the polymers 
irrespective of the values of $M_w$ and $c$, become very clear from the comparisons of these 
figures.  In particular, the normalized birefringence in Fig.\ \ref{samewin} shows an overshoot, 
followed by an undershoot and a second overshoot at exactly similar times for both PS81 and PS2; 
even the magnitude of these undulations are very similar when normalized with the asymptotic 
value, and the first overshoot is higher than the asymptotic value.  In contrast, the normalized 
birefringence for PS82 does not show a clear second overshoot at similar rates of 
flow-deformation; the times at which the overshoot and the following undershoot appear for the 
birefringence of PS82 are, again, very different from that seen with the other two samples, and 
also, the asymptotic value in this case is much higher than the overshoot.  Similarly, the 
orientation angles (Fig.\ \ref{samewichi}) for PS81 and PS2 shows an undershoot behavior, followed 
by an overshoot and a second undershoot (deeper than the first one), each of which appear at a 
slightly later time than the corresponding undulations seen in the birefringence curves.  
These strong correlation in their appearance directly indicates that same microscopic dynamics 
must be responsible for both of these observations.  Thus, the orientation angle traces are very 
similar for these two samples having similar $N_e$ but is very different for PS2, where after 
overcoming the residual glass-birefringence, it almost monotonically decreases to the 
asymptotic value without showing any undershoot behavior.  The similarity in the shape of 
the curves for the measured flow-parameters versus evolution time too are very surprising.  We 
note that the undershoot in flow-parameters (at $t_e \sim 4$~s for PS81 and PS2, and at $t_e 
\sim 12$~s for PS82) is very directly correlated with the first overshoot in the birefringence 
or the first undershoot in the orientation angle (appearing at $t_e \sim 2$~s for PS81 and PS2, 
and at $t_e \sim 9$~s for PS82).  The intensity of these undulations in the flow-parameters 
are also strongly correlated with the intensity of the corresponding undulations in the 
birefringence and orientation angle.  Also, the oscillations are always found to be weaker in 
the case of PS82.  These results, again, very clearly emphasize that the coupled dynamics of 
the flow and the polymer configuration are invariant to changes in the concentration and/or 
molecular weight, provided that the same value of the number of entanglements per chain is 
maintained. 

At these rates of deformation, by comparing the measured and the Newtonian-values of the 
Weissenberg numbers from Table~\ref{table2} for the three polymeric fluids, we note that 
the flow-induced reduction in Weissenberg number or the strain-rate is similar for PS82 and 
PS2.  Thus, we may expect similar values of the asymptotic orientation angle and the 
asymptotic stretch of these two polymers.  In Fig.\ \ref{samewichi}, we indeed see a similar 
long-time value of the orientation angle for these two samples.  As noted earlier, the DEMG 
reptation model predicts that at very high Weissenberg numbers, the polymer chains are extended 
to the maximum chain-extension ratio of $\sqrt{n_t/N_e}$, and the corresponding steady-state 
value of dimensionless birefringence is $\sim n_t/N_e$.  By comparing the asymptotic value of 
the dimensionless birefringence, $\Delta n/CG_N^0$, for PS82 (Fig.\ \ref{n82a}), PS81 (Fig.\ 
\ref{n81b}), and PS2 (Fig.\ \ref{n2b}) with the aforesaid steady-state value, we see that 
compared to the fully extended state, at these similar Weissenberg numbers, the polymer chains 
are stretched to about $10.39 \%$, $5.62 \%$, and $11.04 \%$ for PS82, PS81 and PS2, 
respectively.  The similar extensibility witnessed for PS82 and PS2 is consistent with the 
similar flow-induced reduction of the measured Weissenberg number compared to the Newtonian-value 
for these two solutions.  Between the fluids with the similar high molecular weight polystyrene, 
PS82 and PS81 (refer Table~\ref{table1}), the dilute (and hence the less entangled) solution, 
PS82, has higher molecular weight of the chain-segments between entanglements and hence an 
increased intrinsic extensibility in this case could be expected.  This might be an added reason, 
other than the polymer-modified flow as discussed above, for PS82 showing a higher fractional 
chain-extension compare to PS81 at the similar $Wi_R$.  We note that since the curves for 
$\lambda$ versus $t_e$ (Fig.\ \ref{samewil}) are approximately close to each other for the three 
polystyrene fluids, in order to keep the measured Weissenberg number similar, the velocity-gradient 
curve, \gdot versus $t_e$, in Fig.\ \ref{samewig} (and hence \gdotl versus $t_e$ in 
Fig.\ \ref{samewigl}) is higher for PS2 than that for the other two samples in the same ratio 
as the longest Rouse relaxation time, $\tau_R$, for PS2 is lower than $\tau_R$ for PS81 and PS82. 

\subsection{Characteristic strains for inception of flow}

Following the lead of the work by Geffroy {\em et al.}\cite{enrique1}, we have measured the total 
strain at the peak of the birefringence overshoot for the three entangled polystyrene solutions 
at several different rates of deformation.  The characteristic strain at the peak of the stress 
(or birefringence) overshoot was found to be an useful parameter to determine the transition from 
linear to non-linear viscoelastic behavior of polymers\cite{pearson,enrique1,M&G,koyama}.  It 
was shown in the past that the strain to reach the birefringence overshoot are nearly constant 
at low rates of deformation, but it increases by several orders of magnitude at high rates for both 
simple shear flow\cite{fuller1} and extension-dominated flows\cite{enrique1,koyama}, signifying 
a transition from linear to non-linear viscoelastic behavior.  For the simple shear flow, this 
constant was proposed\cite{M&G} to be about unity, independent of the polymer molecular weight 
and this was experimentally verified\cite{M&G,fuller1} too.  For extensional flows, on the other 
hand, this onset strain was found\cite{enrique1,koyama} to be $\sim 0.7$.  It was proposed\cite{M&G} 
that irrespective of the polymer type and/or molecular weight, the onset of 
the non-linear viscoelastic behavior should begin near $\dot{\gamma}_N \tau_R \sim 1$, which, 
in this particular geometry of the two-roll mill ($\lambda_N = 0.1501$), turns out to be 
$(Wi_R)_N =$ \WiRN $\sim 0.4$, where $(Wi_R)_N$ is the Weissenberg number based on the Newtonian 
values of the velocity-gradient component, \gdotN, the flow-type parameter, \lN, and the longest 
Rouse relaxation time, $\tau_R$.  In previous work with extension-dominated 
flows\cite{enrique1,koyama}, the stain at the overshoot was calculated as the time to reach 
the overshoot times the Newtonian strain-rate, that is $\dot{\gamma}_N \sqrt{\lambda_N} t_p$.  

\begin{figure}
\centerline{\epsfxsize = 8cm \epsfbox{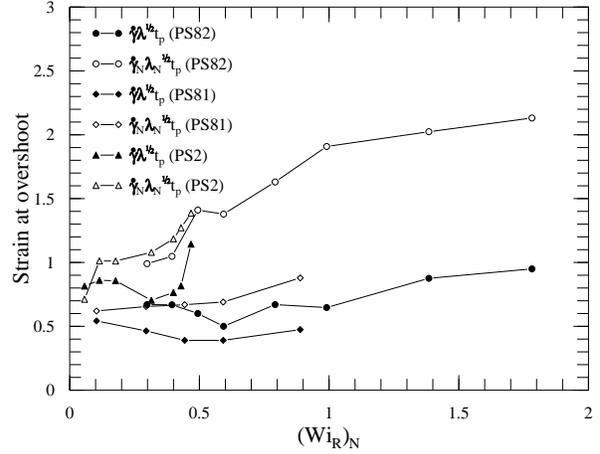}}
\caption{Strain at the overshoot of birefringence against the $(Wi_R)_N = \dot{\gamma}_N 
\lambda_N^{1/2} \tau_R$ for all three polystyrene samples.}
\label{strainn} 
%Fig.~42
\end{figure}

In Fig.\ \ref{strainn}, we have presented our results for the Newtonian as well as the measured 
value of the peak-strain at the overshoot for the three entangled polystyrene fluids as a 
function of $(Wi_R)_N$, in order to simplify the comparisons with the earlier results.  In 
similarity with the earlier observations\cite{enrique1,koyama}, at the lowest rate of 
flow-deformation, the Newtonian-strain at the birefringence overshoot is $\dot{\gamma}_N 
\sqrt{\lambda_N} t_p \sim 0.7$, and its value increases with the deformation 
rate.  Interestingly, this increase with the rate of deformation is similar for samples PS82 and 
PS2, both having $N_e \sim 13$.  At $(Wi_R)_N \sim 0.4$, a modest change in slope in 
$\dot{\gamma}_N \sqrt{\lambda_N} t_p$ is also apparent particularly for these two samples.  In 
contrast, the measured value, $\dot{\gamma} \sqrt{\lambda} t_p$, of the strain at the overshoot 
peak of birefringence, although ``begins'' at a value (viz., 0.5 -- 0.7) that is similar to 0.7, 
it shows an initially decreasing and then increasing trend with the increase of the rate of 
deformation for all three polystyrene samples.  The change in the trend occurs at $(Wi_R)_N$ 
between 0.3 and 0.6 for all three fluids, which is again similar to the proposed value of 0.4.  
The distinct difference between the curves for $\dot{\gamma}_N \sqrt{\lambda_N} t_p$ and 
$\dot{\gamma} \sqrt{\lambda} t_p$ is, again, indicative of the fact that the changes in the 
polymer conformation and the flow-deformation are crucially inter-twined. 

\begin{figure}
\centerline{\epsfxsize = 8cm \epsfbox{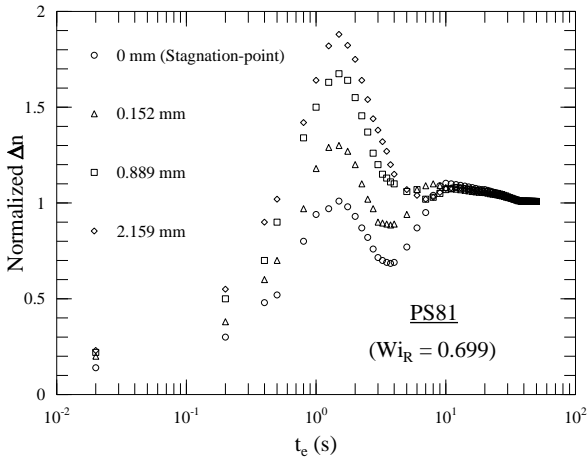}}
\caption{Transient evolution of birefringence, normalized with the corresponding 
asymptotic values, for the same sample, PS81, at the same measured $Wi_R = 0.699$, 
but at several different points across the gap between the stagnation-point and 
the roller surface of the two-roll mill.}
\label{diffptn} 
%Fig.~43
\end{figure}

\subsection{Measurements at different positions within the flow}

It is understandable that the overall flow in a two-roll mill will be very complicated and 
the stagnation region in which we perform our birefringence as well as flow measurements is a 
very small and special region of the entire flow where the flow can be approximated by the 
uniform and homogeneous form given by Eqn.\ (\ref{v}), for both Newtonian and viscoelastic fluids.
Hence, to try to measure and interpret the polymer conformation, the flow kinematics and/or 
their coupling in any region of the flow other than the stagnation would be an extremely 
difficult experimental task.  Earlier numerical simulation results (using the Chilcott-Rallison 
model\cite{CR} for dilute polymeric solutions\cite{singh}, and the vector model\cite{johan} for 
the entangled polymer solutions) for the present configuration of the two-roll mill (i.e., with 
\lN = 0.1501), had shown that at a constant rate of imposed flow-deformation, the extensional 
character of the flow (i.e., the value of $\lambda$) as well as the rate of strain, \gdotsql, 
reduce compared to their Newtonian-value, \lN, and \gdotsqlN, respectively, (or the flow becomes 
more shear-like) in the region close to the roller surface.  In fact for the entangled 
fluids\cite{johan}, the rate of strain in the gap between the rollers smoothly changes its value 
from a minimum at the stagnation-point to a maximum at the roller surface.  The difference 
between the minimum and the maximum could be as high as three to four times depending on the 
concentration of the sample.  

\begin{figure}
\centerline{\epsfxsize = 8cm \epsfbox{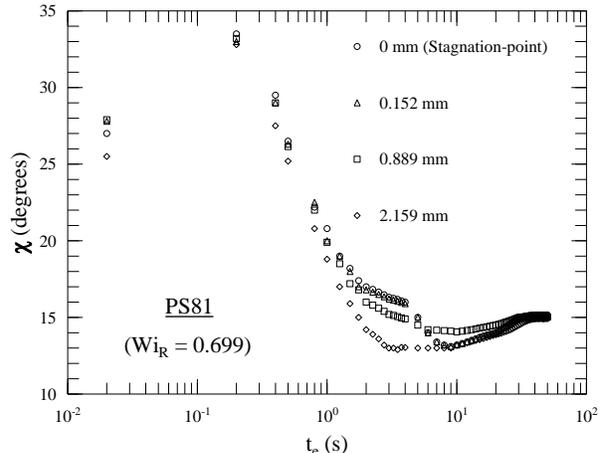}}
\caption{Same as Fig.~43, but for inception of orientation angle, $\chi$.}
\label{diffptchi} 
%Fig.~44
\end{figure}

We have attempted to measure the birefringence and the flow-parameters 
at different points in the two-roll mill flow from the stagnation-point to the roller-surface 
when the two-roll mill is orientated along the $y$ axis in Fig.\ \ref{two-roller2}, i.e., 
when $\phi = 90^\circ$.  The flow measurements are also performed for 
the other orientation of the two-roll mill, i.e., when it is oriented along $x$ axis ($\phi = 
0^\circ$).  It is intrinsic to our set-up that the two-roll mill can be rotated only about the 
central stagnation-point.  This restricts us to measure only one velocity-gradient component at 
different points along the $x$ and $y$ axes when the flow-cell is orientated along the same 
directions.  At the stagnation-point these two velocity-gradient components are just \gdot and 
\gdotl, respectively.  In these attempts, we have assumed that the flow at different positions 
along these two axes, to a first approximation, can still be described by Eqn.\ (\ref{v}), so 
that the process of measuring the velocity-gradient parameters via dynamic light scattering is 
still valid.  Although this has not been rigorously verified, pending the availability of a more 
detailed calculation for measuring the velocity-gradient parameters at different points in the 
flow, we can still use the same DLS technique to get an approximate picture of the flow kinematics.  
The distance from the stagnation-point to the surface of the rollers for the present configuration 
of the two-roll mill is 4.22 mm, but the gear assembly attached in front of the two-roll mill 
obstructs the beam to go through the sample-cell beyond a distance of 1.3 mm from the 
stagnation-point when the flow-cell is orientated along $y$ axis.  This restriction does not 
apply when the two-roll mill is rotated along $x$ axis.  

\begin{figure}
\centerline{\epsfxsize = 8cm \epsfbox{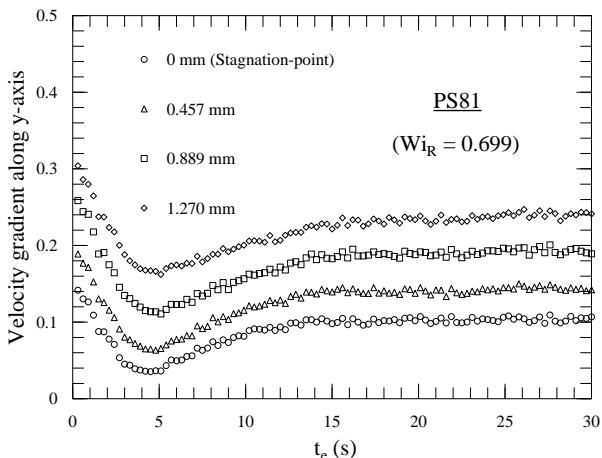}}
\caption{Transient evolution of the velocity-gradient component at several 
different points along $y$ axis of Fig.~1 for sample, PS81, at the measured 
Weissenberg number, $Wi_R = 0.699$, and the two-roll mill orientation, $\phi = 90^\circ$.}
\label{diffpty} 
%Fig.~45
\end{figure}

Fig.\ \ref{diffptn} and \ref{diffptchi} 
show the temporal-evolution of the birefringence (scaled with the corresponding asymptotic value), 
and the orientation angle at four different points along the $y$ axis on inception of flow to a 
measured Weissenberg number of 0.699 for the polystyrene fluid PS81.  The birefringence at the 
stagnation-point shows a very distinct overshoot and undershoot behavior before reaching the asymptotic 
value, which is higher than the peak overshoot.  This is a characteristic feature shown by PS81 
when subjected to extension-dominated flows, as we have noticed before in Fig.\ \ref{n81b}.  As one 
approaches closer to the roller-surface from the stagnation-point, the strength of the overshoot 
increases but the undershoot diminishes so that we finally obtain a birefringence curve with a 
single peak and a slight undershoot measured at 2.159 mm away from the stagnation-point towards 
the surface of the rollers.  Before the reaching the asymptotic value, the temporal-dependence 
of the orientation angle also exhibits a transition from a distinct undershoot behavior at 
stagnation-point and in the region close to it, characteristic of an extension-dominated flow, 
to a smaller and wider undershoot behavior quite away from the stagnation-point, characteristic 
of a more shear-like flow.  This confirms the theoretical findings\cite{johan,singh}, as 
discussed above.  The different curves for the orientation angle versus evolution time exhibit 
most differences from each other in the time-period $t_e$ = 0.7 s -- 30 s, which is shifted in 
time from the time-period $t_e$ = 0.2 s -- 10 s, over which the corresponding birefringence 
curves show most differences from each other.  This, again, is consistent with our earlier 
observations.  It is important to note from Fig.\ \ref{diffptchi}, that at the same measured 
$Wi_R$ the polymer chains are asymptotically orientated to the same angle, $\chi_\infty \sim 
15^\circ$, at different points studied within the gap between the rollers. 

\begin{figure}
\centerline{\epsfxsize = 8cm \epsfbox{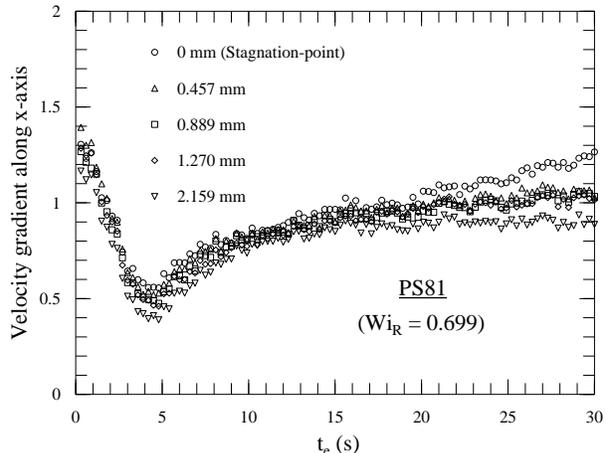}}
\caption{Same as Fig.~45, but along $x$ axis of Fig.~1 and ~2, and $\phi = 0^\circ$.}
\label{diffptx} 
%Fig.~46
\end{figure}

The time-dependent velocity-gradient component along $y$ axis for the inception of the same flow 
is depicted in Fig.\ \ref{diffpty}.  As noted before, at the stagnation-point, this parameter 
is simply \gdotl, which already shows the impact of the polymer induced flow-modification even at 
the first point of our measurement, i.e., as early as $t_e = 0.3$~s, where its measured value 
is shown to be lower than the Newtonian-value of 0.2.  The temporal behavior of this component 
at different positions along $y$ axis, for $\phi = 90^\circ$, is surprisingly similar in shape to 
that of $\dot{\gamma}\lambda$.  In particular, all of them ``begin'' with an initial higher 
value, pass smoothly through a minimum and again increase to reach the asymptotic value (which 
is lower than the initial value) as early as $t_e \sim 15$~s.  This component of the 
velocity-gradient tensor increases in magnitude when approached towards the surface of the rollers 
from the stagnation-point.  In sharp contrast to this, the component along $x$ axis, when $\phi = 
0^\circ$, does not change much between the two rollers.  This is clearly shown in Fig.\ 
\ref{diffptx}, where the different curves almost overlap on to each other until about $t_e = 15$~s, 
beyond which the asymptotic value is reached faster in time in the region closer to the 
roller-surface, compared to that in the stagnation-point, where even at $t_e = 30$~s, this 
component of velocity-gradient still shows an increasing trend.  A closer look at the differences 
among these curves indicates that there is a decreasing trend in the value of this velocity-gradient 
component away from the stagnation-point at each instant of time during the flow evolution.  This 
behavior is opposite to what is seen in Fig.\ \ref{diffpty}.  Interestingly, for all the curves 
shown in Figs.\ \ref{diffpty} and 
\ref{diffptx}, the time at which the minimum of the undershoot is reached is almost constant, 
$t_e \sim 4$~s.  Similarly, the peak of the birefringence for all the curves occurs at $t_e 
\sim 2$~s, and these to times are strongly correlated to each other.  As we have noted before, 
these times are direct functions of the Weissenberg number of the flow and the molecular 
characteristics of the polymer, both of which are held fixed in this case.  These observations 
are qualitatively in a very close agreement with that found using the vector model\cite{vector} 
by Remmelgas {\em et al.}\cite{johan} for the start-up of two-roll mill flows for entangled 
polymeric solutions.  It is important to note that given the inherent difficulty of our set-up 
to measure the other velocity-gradient components corresponding to Figs.\ \ref{diffpty} and 
\ref{diffptx}, a critical comparison to the findings of the above model on the flow in the gap 
between the rollers, discussed above, is impossible.  Nevertheless, it is clear that it would 
be extremely worthwhile to calculate the predictions from this model for these experimentally 
measurable parameters and that should be compared to the present findings for a further insight 
on the dynamics of the polymer microstructure and the velocity-field.  

\section{Summary and Conclusions} 

We have presented our results on the transient evolution of birefringence and orientation angle 
for a series of entangled polymer solutions subjected to the start-up flow at several Weissenberg 
numbers.  An extensive study of these results along with the corresponding velocity-gradients for 
each Weissenberg number and for each sample, measured using a dynamic light scattering technique, is 
presented for the first-time for such extension-dominated time-dependent flows.  As expected, 
there is a very significant flow-modification geared by the orientation and stretching of the 
polymer molecules.  The measured flow kinematics show that the effect of the polymers on the 
flow in the immediate vicinity of the stagnation-point is to decrease its extensional strength 
(and the rate of strain) relative to the velocity-field for a Newtonian fluid.  This drastically 
changes the value of the steady-state $Wi_R$ measured at the stagnation-point, which we have 
used to identify each transient experiment, from that imposed on the polymeric fluids.  Similar 
results of polymer induced reduction of strain-rate at the stagnation-point of a two-roll mill 
was also obtained experimentally for dilute polymer solutions\cite{graham,ng}.  This observation 
was qualitatively interpreted\cite{graham,singh}, using a dumbbell model\cite{CR}, from the fact 
that dilute solution of high molecular weight, linear, flexible polymers strongly strain-rate 
harden in extension-dominated flow\cite{orr}.  In contrast, entangled polymeric fluids 
strain-rate soften over a considerable range of strain-rates in such flows and hence the 
experimentally observed similar behavior for these very different polymer concentrations seems 
to be quite remarkable.  However, recent numerical studies by Remmelgas {\em et al.}\cite{johan}, 
using a vector approximation of the DEMG model in a two-roll mill, have indeed predicted a 
decrease in the strain-rate at the stagnation-point for entangled polymeric fluids, in clear 
agreement with our experiments.  They interpreted that this decrease results from the effect of 
a shear-thinning viscosity in the flow near the rollers which makes momentum transfer to the 
region between the rollers less efficient.   This prediction is specific to the 
extension-dominated flows generated in the two-roll mill only.

We observed a very distinct coupling between the polymer conformational dynamics and the changes 
in the flow-field.  The results for the time-dependent flows presented 
in this paper, the corresponding steady-state results in Part I\cite{pst1}, as well as some other 
studies from our laboratory\cite{dmitry} unequivocally emphasizes that the in the flow 
regime away from chain-stretching, these coupled dynamics are invariant to changes in polymer 
concentration and molecular weight, as long as the number of entanglements per chain is held 
fixed.  On the inception of flows with similar Weissenberg numbers, an increase in the number of 
entanglements per chain was accompanied by increased overshoot and undershoot in the 
birefringence and by a slower response time, in addition to a decreased steady-state stretching 
of the polymer chains.  There is an associated undershoot in the orientation angle data at high 
rates of deformation which is shifted to longer times compared to the first overshoot in the 
birefringence.  The polymer induced flow behavior is very complicated.  The variation of the 
magnitude of the flow-parameters over the total evolution time is very pronounced for the fluid 
with lowest $N_e$, in contrast to a weak effect found for the other two samples having similar 
value for $N_e$.  There is a transition in the behavior of the polymer anisotropy as well as the 
coupled flow dynamics at some critical Weissenberg number specific to each system.  The 
velocity-gradient component, \gdotl, and hence the flow-type parameter, $\lambda$, are more 
sensitive to the impact of evolution of the polymer microstructure, than the other velocity-gradient 
component, \gdot.  The undershoot in the measured flow-parameters seem to be strongly correlated 
with the first overshoot in the birefringence and hence the first undershoot in the orientation 
angle data.  The Newtonian-strain at the overshoot of birefringence for the polystyrene solutions 
shows qualitatively similar behavior to that found earlier\cite{enrique1,koyama} for the 
extension-dominated flows.  The behavior for the measured strain, in contrast, differs from that 
because of the strong modification observed in the flow from the corresponding Newtonian-value.  
In similarity with that seen with dilute polymeric solutions\cite{graham,ng}, the flow with the 
entangled solutions too reduces its extensional character and becomes more viscometric or 
shear-like away from the stagnation-point although the underlying mechanism in these two cases 
was shown, in a recent theoretical work\cite{johan}, to be very different from each other.  The 
measured velocity-gradient shows different behaviors in the gap between the rollers depending on 
the orientation of the two-roll mill in the plane of the flow.  At small Weissenberg numbers the 
flow-type parameters extracted from the dynamic light scattering results have values that are 
consistently higher than that would be expected for a Newtonian fluid.  Our steady-state 
results\cite{pst1} on the same polymeric fluids also support this finding.  This is the first-time 
observation of such a behavior of the flow-type parameter for entangled polymer solutions in 
the range of small Weissenberg numbers for a strong, extension-dominated flow.  The numerical 
simulations by Remmelgas {\em et al.}\cite{johan} have clearly shown, in direct similarity 
with our experimental results, that the flow-type parameter for entangled polymer solutions near 
the stagnation-point for this particular configuration of the two-roll mill does increase relative 
to the Newtonian-values, at small Weissenberg numbers.  This point certainly deserve further 
careful experimental and theoretical studies, and ongoing experiments in our laboratory are 
intended to resolve this issue.  Nevertheless, the results presented here provides an unique 
basis to crucially test the reptation based DEMG models with chain-stretching included.  This 
can be done by using the experimentally measured time-dependent velocity-gradient as an input 
to the model and then comparing its prediction for the time-dependent birefringence and 
orientation angle to that obtained experimentally.  Alternatively, for the range of parameters 
involved in this study, one can numerically simulate both the birefringence and velocity-field, 
using the computationally tractable vector approximation of the DEMG model\cite{vector}, which 
could then be compared with the present experiments.  In contrast to the first method, the 
second one simulates the predictions of both the effect of flow on polymers and the effect of 
polymers on flow.  Work is underway in these directions. 

\acknowledgments

We thank Johan Remmelgas and James P. Oberhauser for helpful discussions.

%\widetext

\narrowtext 

\begin{table}
\caption{The characteristic parameters for the three polystyrene samples.}
\begin{tabular}{cccccccc} 
Sample & $M_w$ & $M_w/M_n$ & $c$ & $\eta_0$ & $N_e$ & $\tau_R$ & $n_t$ \\ 
& ($\times 10^6$) && ($\frac{\hbox{g}}{\hbox{cc}}$) & (P) && (s) \\ \tableline
PS81 & 8.42 & 1.17 & 0.0396 & 7500 & $\sim 13$ & 2.25 & 8420 \\
PS82 & 8.42 & 1.17 & 0.0262 & 2700 & $\sim 7$ & 3.01 & 8420 \\
PS2 & 2.89 & 1.09 & 0.0867 & 19000 & $\sim 13$ & 0.56 & 2890 \\
\end{tabular}
\label{table1}  
\end{table}

\begin{table}
\caption{The Weissenberg numbers [measured, $Wi_R = \dot{\gamma} \lambda^{1/2} \tau_R$, 
and the Newtonian,  $(Wi_R)_N = \dot{\gamma}_N \lambda_N^{1/2} \tau_R$] used for the three 
polystyrene samples and the percentage flow-modification.}
\begin{tabular}{ccccccccc} 
\multicolumn{3}{c}{PS82} & \multicolumn{3}{c}{PS81} & \multicolumn{3}{c}{PS2} \\ \tableline
$(Wi_R)_N$ & $Wi_R$ & \% & $(Wi_R)_N$ & $Wi_R$ & \% & $(Wi_R)_N$ & $Wi_R$ & \% \\
0.099 & 0.078 & 21.33 & 0.052 & 0.004 & 92.04 & 0.031 & 0.013 & 59.09 \\
0.199 & 0.144 & 27.56 & 0.058 & 0.006 & 90.58 & 0.057 & 0.027 & 51.89 \\ 
0.298 & 0.199 & 33.26 & 0.081 & 0.011 & 86.83 & 0.114 & 0.056 & 51.21 \\
0.394 & 0.245 & 37.88 & 0.103 & 0.017 & 83.45 & 0.177 & 0.075 & 57.66 \\
0.494 & 0.284 & 42.42 & 0.295 & 0.103 & 65.19 & 0.315 & 0.151 & 52.10 \\
0.593 & 0.319 & 46.20 & 0.443 & 0.181 & 59.15 & 0.400 & 0.215 & 46.32 \\
0.792 & 0.384 & 51.49 & 0.592 & 0.266 & 55.04 & 0.429 & 0.236 & 45.00 \\
0.990 & 0.455 & 54.05 & 0.889 & 0.470 & 47.13 & 0.468 & 0.263 & 43.82 \\
1.385 & 0.636 & 54.07 & 1.183 & 0.699 & 40.95 & & & \\
1.782 & 0.843 & 52.70 & & & & & \\
\end{tabular}
\label{table2}  
\end{table} 

\end{document}